\documentclass{aa}

\usepackage{natbib}
\usepackage[normalem]{ulem}

\bibpunct{(}{)}{;}{a}{}{,} 
\usepackage{color}
\usepackage{graphicx}
\usepackage{amssymb}

\begin{document}

\renewcommand{\vec}[1]{\bmath{#1}}
\newcommand{\sign}{\textrm{sign}}
\newcommand{\pd}[2]{\frac{\partial #1}{\partial #2}}
\newcommand{\DS}{\displaystyle}
\newcommand{\HALF}{\frac{1}{2}}
\newcommand{\mathi}{\rm i}

\renewcommand{\vec}[1]{\mathbf{#1}}
\newcommand{\hvec}[1]{\hat{\mathbf{#1}}}
\newcommand{\av}[1]{\left<#1\right>}
\newcommand{\red}[1]{\color{red} #1 \color{black}}
\newcommand{\blue}[1]{\color{blue} #1 \color{black}}

\title{The different flavors of extragalactic jets: \\ Magnetized relativistic flows}

\author{}
\author{P. Rossi\inst{1}, G. Bodo\inst{1}, S. Massaglia\inst{2}, A. Capetti\inst{1} }

\authorrunning{P.Rossi et al.}
\titlerunning{Deceleration of relativistic jets}

\institute{INAF/Osservatorio Astrofisico di Torino, via Osservatorio 20, 10025 Pino Torinese, Italy  \and Dipartimento di Fisica, Universit\`a degli Studi di Torino, via
  Pietro Giuria 1, 10125 Torino, Italy}

\date{Received ?? / Accepted ??}


\label{firstpage}

\abstract
{We perform three-dimensional numerical simulations of magnetized relativistic jets propagating in a uniform density environment in order to study the effect of the entrainment and the consequent deceleration, extending a previous work in which magnetic effects were not present. As in previous papers, our aim is to understand the connection between the jet properties and the resulting Fanaroff-Riley classification. We consider jets with different low densities, and therefore low power, and different magnetizations. We find that lower magnetization jets effectively decelerate to sub-relativistic velocities and may then result in an FR~I morphology on larger scales. At the opposite, in the higher magnetization cases, the entrainment and consequent deceleration are substantially reduced.
}

\maketitle

\section{Introduction}
%
%
%

The traditional Fanaroff-Riley classification \citep{FR74} divides extragalactic radio-sources in two groups, it is based on morphological properties, but it is also reflected in radio luminosity. Fanaroff-Riley I (FR~I) are low radio power sources and are characterized by jet-dominated emission that smoothly extends into the intracluster medium, where large-scale plumes or  tails of diffused emission are observed. The second class, Fanaroff-Riley II (FR~II), is characterized by higher radio power and the sources in this class show the maximum brightness in the hot spots at the jet termination. Recently a third  class of  radio-galaxies has been introduced, representing the majority of the local radio-loud AGN population. Because they lack the prominent extended radio structure characteristic of the other Fanaroff-Riley classes they were called FR~0 \citep{Baldi09, Baldi10b, Sadler14, Baldi15}. There is however a minority of FR~0 sources that show evidences for small-scale jets, usually limited to sizes of a few kpc \citep{Baldi19}. This suggests that jets in this class are not even able to escape from the galaxy core. 
There are observational evidences that, both in FR~I and in FR~II, the jets at their base (at the parsec scale) are relativistic, with similar Lorentz factors \citep{gg01, Celotti08}. The situation is less clear for FR~0 sources, \citet{giovannini23}, from VLBI observations, conclude that jets in FR~0 can be, at most,  mildly relativistic, however \citet{Massaro20}  suggest that FR~0 could represent the misaligned counterpart of a significant fraction of BL Lac objects and therefore they also should be associated to relativistic jets. This apparent contradiction could be solved by considering a jet with a transverse velocity structure, with a relativistic spine surrounded by a non relativistic flow. It is clear that in FR~I jets and, possibly also in FR~0 jets, a deceleration to sub-relativistic velocities must occur between the inner region (at the parsec scale) and the kiloparsec scale, where FR~I show their plume-like morphology \citep{laing02a, Laing14} and FR~0 disappear.  \citet{massaglia16} (hereinafter Mas16) performed high-resolution three-dimensional simulations of low Mach number jets, showing the onset of turbulence and the formation of structures very similar to that observed in FR~I sources \citep[see also][]{westhuizen19}. However, in this study, the jets are injected already at non-relativistic velocities and one has to assume that deceleration has already occurred.

Deceleration occurs most likely through the progressive turbulent entrainment of external material \citep[see e.g.][]{DeYoung93, DeYoung05, Bicknell84, Bicknell86, Bicknell94, wang09}, as a consequence of the development of jet instabilities, in particular Kelvin-Helmholtz \citep[see e.g.][]{peru07, rossi08, rossi20}, but also of other kinds. The role of current-driven instabilities  has been considered, for example,  by \citet{Tchekhovskoy16}, \citet{massaglia19, massaglia22} and \citet{mukherjee20}. Moreover,  it has been recently pointed out that recollimation shocks may trigger instabilities and possible entrainment, in particular \citet{gourgouliatos18a, gourgouliatos18b} and \citet{bromberg23} discussed the development of the relativistic centrifugal instabilities and \citet{costa23} consider Kelvin-Helmholtz instabilities. In addition, we must also mention a second option, that may be complementary to turbulent entrainment, related to  mass loading by stellar mass loss \citep{komissarov94, bowman96, laing02b, perucho14a, perucho14b,  perucho20, perucho21}.

In this paper we extend the work  by \citet{rossi08} and \citet{rossi20} (hereinafter Paper I and Paper II). They performed numerical simulations of the propagation of relativistic unmagnetized jets, with different density ratios between the jet and the ambient medium, studying in detail the entrainment and deceleration process.  They found that lighter jets are more prone to deceleration. In particular, in Paper II, we followed the jet propagation up to a distance of about 600 initial jet radii, showing how jets, with a density ratio of $10^{-4}$, are able to reach a quasi-steady configuration, where the Lorentz factor decreases from 10, at the injection point, to about 2, close to the jet head. On the contrary, denser jets can propagate in an almost undisturbed way.  The main limitations of the analysis presented in Papers I and II is the absence of magnetic field, our aim in the present paper is to overcome this limitation. 

 In this paper we concentrate our analysis on the role of the density ratio and of the field strength in determining the jet deceleration process and we will show that these parameters play a very important role in this process. Of course, there are also other aspects are likely to play a relevant role on the jet stability and evolution, such as i) different magnetic field configurations, ii) different initial Lorentz factors, iii) inhomogeneities in the ambient medium. However the parameter space is very wide and one has to proceed in a step-by-step manner.

In Sect. \ref{sec:numsetup} we describe the set of equations to be solved, the numerical method, the initial and boundary conditions and the parameters of the simulations. In Sect. \ref{sec:results} we present our results and in Sect. \ref{sec:discussion} we give discussion and conclusions.

\section{Numerical setup}
\label{sec:numsetup}
%
%
%
%
%
%
Numerical simulations are carried out by solving the equa-
tions of relativistic ideal magneto-hydrodynamics in 3D. They are expressed by
\begin{equation}\label{eq:continuity}
    \partial_t \left( \gamma \rho \right) + \nabla \cdot \left( \gamma \rho \mathbf{v }\right) =  0 \,,
\end{equation}

\begin{equation}
    \partial_t \left( \gamma^2 \rho h \mathbf{v} + \mathbf{E} \times \mathbf{B}   \right) +
    \nonumber
\end{equation}
\begin{equation}\label{eq:momentum}
    + \nabla \cdot \left[] \gamma^2 \rho h \mathbf{v} \mathbf{v} - \mathbf{E} \mathbf{E} - \mathbf{B} \mathbf{B} + \left( p + u_{em}\right) \mathbf{I}\right] = 0 \,,
\end{equation}

\begin{equation}\label{eq:energy}
    \partial_t \left( \gamma^2 \rho h -p + u_{em} \right) + \nabla \cdot \left( \gamma^2 w \mathbf{v} + \mathbf{E} \times \mathbf{B} \right) = 0 \,,
\end{equation}

\begin{equation}\label{eq:induction}
    \partial_t \mathbf{B} + \nabla \times \mathbf{E} = 0 \,,
\end{equation}
where the electric field $\mathbf{E}$ is provided by the ideal condition $\mathbf{E} + \mathbf{v} \times \mathbf{B} = 0$. In this set of equations,  $\rho$ is the proper density,  $\mathbf{v}$ is the velocity three-vector (in units of the light speed $c$), $\mathbf{B}$ is the laboratory magnetic field (in units of $\sqrt{4 \pi}$), $\gamma = 1/ \sqrt{1-v^2}$ is the Lorentz factor,  $h$ is the relativistic specific enthalpy, $u_{em} = (E^2+B^2)/2$ and $\mathbf{I}$ is the unit $3 \times 3$ tensor. The system of equations (\ref{eq:continuity}-\ref{eq:induction}) is completed by specifying an equation of state relating $h$, $p$ and $\rho$. Following \citet{Mignone05}, we adopted the Taub-Mathews (TM) equation of state, with the prescription:
\begin{equation}\label{eq:enthalpy}
    h = \frac{5}{2} \frac{p}{\rho} + \sqrt{1 + \frac{9}{4} \frac{p^2}{\rho^2}} \,,
\end{equation}
which closely reproduces the thermodynamics of a Synge gas for a single species relativistic perfect fluid (notice that we set $c=1$, since the light speed is our unit of velocity).

In addition to the system (\ref{eq:continuity}-\ref{eq:induction}), we also integrate the equation
\begin{equation}
    \partial_t \left( \gamma \rho f  \right) + \nabla \cdot \left( \gamma \rho f \mathbf{v}\right) = 0 \,,
\end{equation}
which describes the evolution of a passive tracer $f$, set to unity for the injected jet material and to zero for the ambient medium. In this way we can follow the evolution of the spatial distribution of the jet material and its mixing with the ambient medium.

Simulations are carried out in a Cartesian domain with coordinates in the range $x \in \left[ -L_x/2, L_x/2 \right]$, $y \in \left[ 0, L_y \right]$ and $z \in \left[ -L_z/2, L_z/2 \right]$. At $t=0$ the domain is filled with a perfect fluid of uniform density and pressure, representing the external medium initially at rest. In addition, we assume the external medium to be initially unmagnetized. The assumption of constant density is consistent with the fact that we focus on the jet propagation inside the galaxy core.

A cylindrical inflow of velocity $v_j$, proper density $\rho_j$ and a purely azimuthal magnetic field is constantly fed into the domain at  $y = 0$,  along the $y$ direction. The jet is characterized by its radius $r_j$, the Lorentz factor $\gamma_j$, the ratio $\eta_j$ between its proper density $\rho_j$ and the external density and the Mach number defined as $M_j = v_j/c_{s0}$, where $c_{s0}$ is the sound speed on the jet axis. The sound speed $c_{s0}$ is given by the expression valid for the TM equation of state \citep{Mignone05}
\begin{equation}
    c_{s0} = \frac{p_0}{3 \rho_j h_0} \frac{5 \rho_j h_0 - p_0}{\rho_j h_0 - p_0} \,,
\end{equation}
where $p_0$ and $h_0$ are respectively the pressure and the specific enthalpy computed on the jet axis. The specific enthalpy can be derived through Eq. \ref{eq:enthalpy}.  The injected magnetic field has a configuration corresponding to a constant current density inside $r = a$ (we take $a=0.75 r_j$) and zero outside:
\begin{equation}
    B_\varphi = 
    \begin{cases}
    -B_m r/a \,, & \text{if} \; r \leq a \\
    -B_m a/r            & \text{if} \; r > a \,,
    \end{cases}
\end{equation}
where $B_m$ is the maximum magnetic field strength reached at $r=a$.  The return current during the jet evolution follows the backflow and the shocked ambient medium similarly to what is seen  in \citet{leismann05}. Inside the jet $(r < r_j$), we assume total pressure equilibrium, therefore the thermal pressure inside the jet is given by
\begin{equation}
    p_j(r) = p_0 - B_m^2 {\rm min} \left(\frac{r^2}{a^2}, 1 \right) \,.
\end{equation}
The value of $p_0$, as discussed above, is related to the jet Mach number $M_j$, while the value of $B_m$, in the jet rest frame,  can be obtained through the magnetization parameter $\sigma_j =B_m^2/\rho_j h_0$. The external medium can be characterized  the parameter 
\begin{equation}
    \Pi = \frac{p_a}{\rho_a}  \,,
\end{equation}
where $p_a$ and $\rho_a$ are respectively the pressure and density of the external medium.

The equations are non-dimensionalised by using $c$ as the unit of velocity (as already discussed), the jet radius $r_j$ as the unit of length and the ambient density $\rho_a$ as the unit of density. To specify completely the problem we therefore need the following non-dimensional parameters, introduced above: the jet Lorentz factor $\gamma_j$, the density ratio $\eta_j$, the jet Mach number $M_j$, the jet magnetization $\sigma_j$, and the non-dimensional ambient pressure $\Pi$. In our analysis we kept fixed the Lorentz factor $\gamma_j = 10$, the Mach number $M_j = 2.2$ and $\Pi=5 \times 10^{-7}$ (corresponding to a temperature of the ambient medium of $5 \times 10^6$K).  We varied instead the density ratio $\eta_j$ and the magnetization $\sigma_j$ for a total of four simulations. All the jets results to be over-pressured with respect to the ambient medium. The parameters of the simulations are reported in Table \ref{tab:cases}, where we also report the pressure contrast $K$ between the jet and the ambient, the relativistic specific enthalpy in the jet  and the jet power discussed below.

\begin{table}[!ht] \label{tab:cases} 
\centering

\begin{tabular}{cccccc}\hline                                   
Case    & $\eta_j$    &  $\sigma_j$ & $K$ & $h_0$ & $P_j$ [erg s$^{-1}$]   \\
\hline\hline\noalign{\medskip}
A      & $10^{-4}$        & $10^{-2}$     & $20$  & 1.6 & $1.96 \times 10^{43}$ \\
B      & $10^{-4}$        & $10^{-1}$     & $20$   & 1.6 &$2.04 \times 10^{43}$\\
C      & $10^{-5}$        & $10^{-2}$     & $2$ & 1.6 &  $1.96 \times 10^{42}$\\  
D      & $10^{-5}$        & $10^{-1}$     & $2$  &1.6  & $2.04 \times 10^{42}$ \\   
 \noalign{\medskip}
\hline
\medskip
\end{tabular}
\caption[]{Parameter set used in the numerical simulation model. The second 
column refers to density ratio $\eta_j$, the third to the magnetization $\sigma_j$, the fourth  gives the derived pressure ratio,  the fifth gives the value of the relativistic specific enthalpy  in the jet  and the last column to the jet power evaluated according to Eq. (\ref{eq:jetpower})}
\label{parset}
\end{table}

Numerical integration of the system (\ref{eq:continuity}-\ref{eq:induction}) is performed using the ideal relativistic magnetohydrodynamics module in the PLUTO code \citep{PLUTO}. For the present application we employed linear reconstruction, second order Runge-Kutta time stepping and the HLLD Riemann solver \citep{mignone09}. Constrained transport \citep{balsara99, londrillo04} is used to maintain the condition $\nabla \cdot \mathbf{B} = 0$. For the physical domain extension we choose $L_x = 300$, $L_y = 600$ and $L_z = 300$ and we cover this domain with a grid of $350 \times 2400 \times 350$ grid points, which are uniformly spaced in the region around the jet axis ($-15 < x <15$, $-15 < z <15$) and geometrically stretched outside.  The cells are elongated in the jet directions and we have a resolution of 8 points on the jet radius in the $x$ and $z$ directions and of 4 points on the jet radius in the $y$ direction. The boundary conditions are outflow on all the boundaries except for $y=0$, where, for $r<r_j$, we have inflow conditions and, for $r>r_j$ we have reflective conditions.

\section{Results}
\label{sec:results}
%
%

\begin{figure}
    \centering
    \includegraphics[width=\columnwidth]{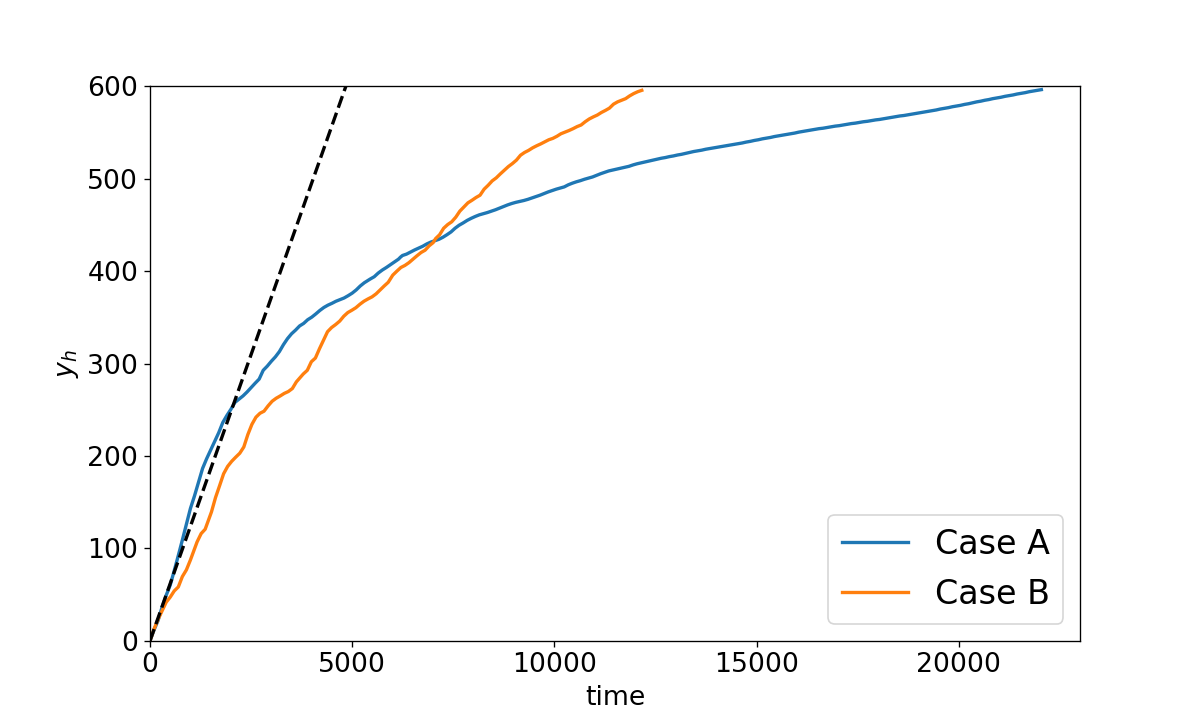}
    \includegraphics[width=\columnwidth]{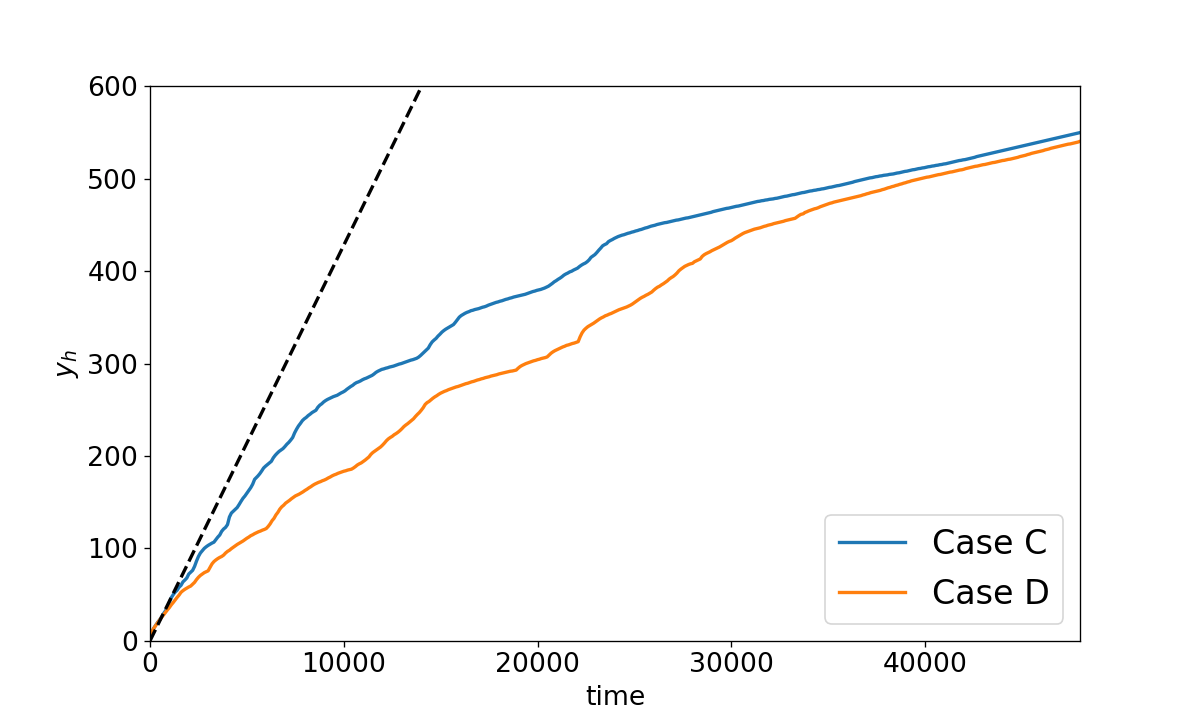} 

    \caption{Position of the jet head as a function of time, the top panel shows cases A and B, the bottom panel shows cases C and D.}
    \label{fig:yhead}
\end{figure} 

\begin{figure*}
    \centering
    \includegraphics[width=\columnwidth]{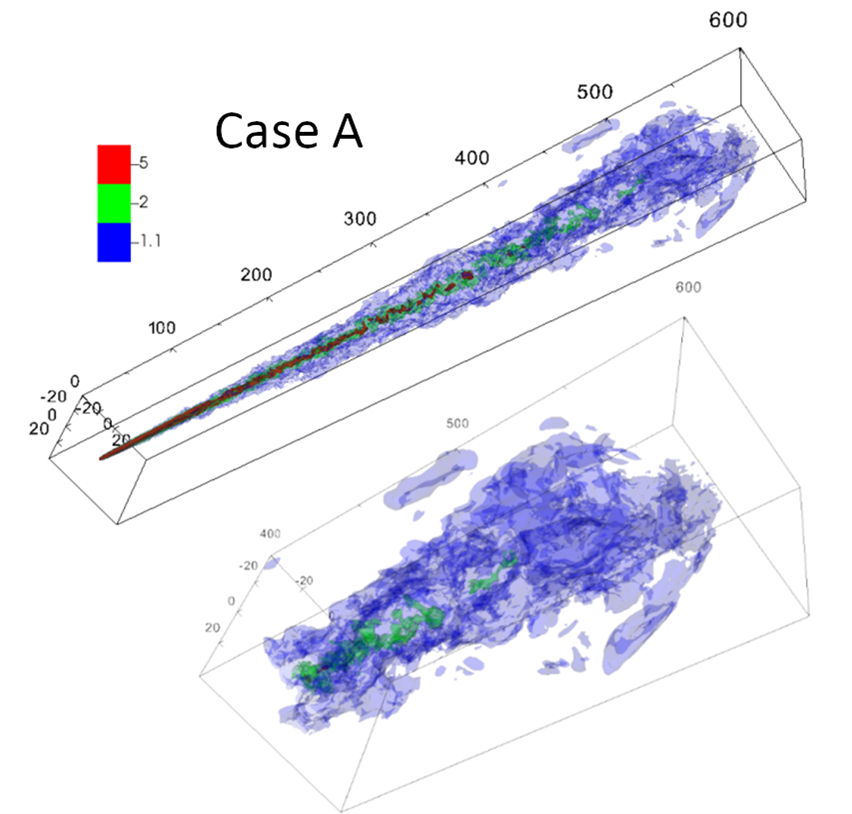}
   \includegraphics[width=\columnwidth]{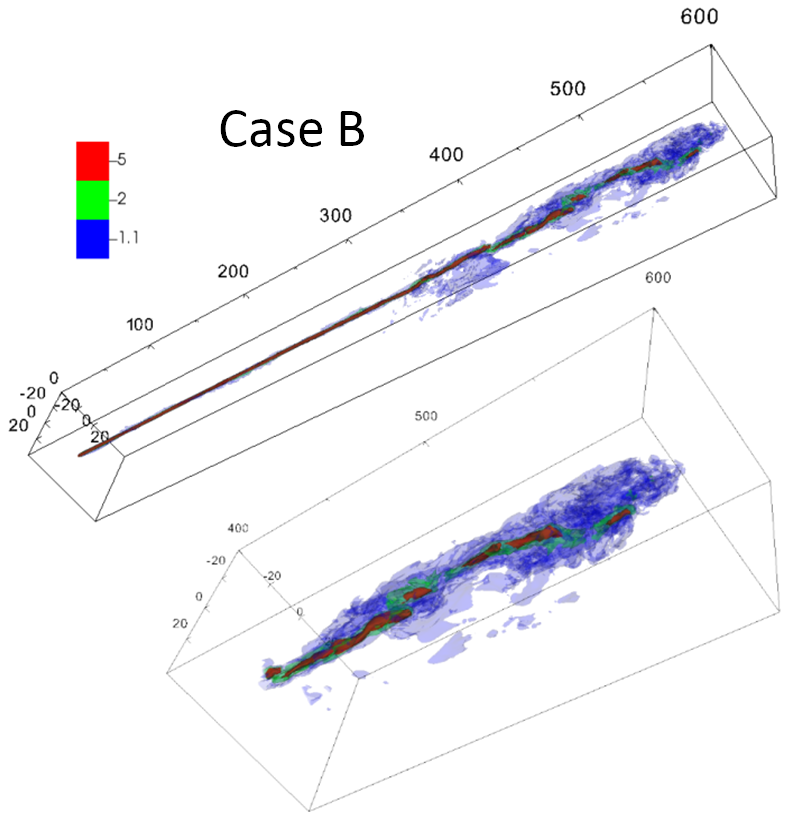} \\
    \includegraphics[width=\columnwidth]{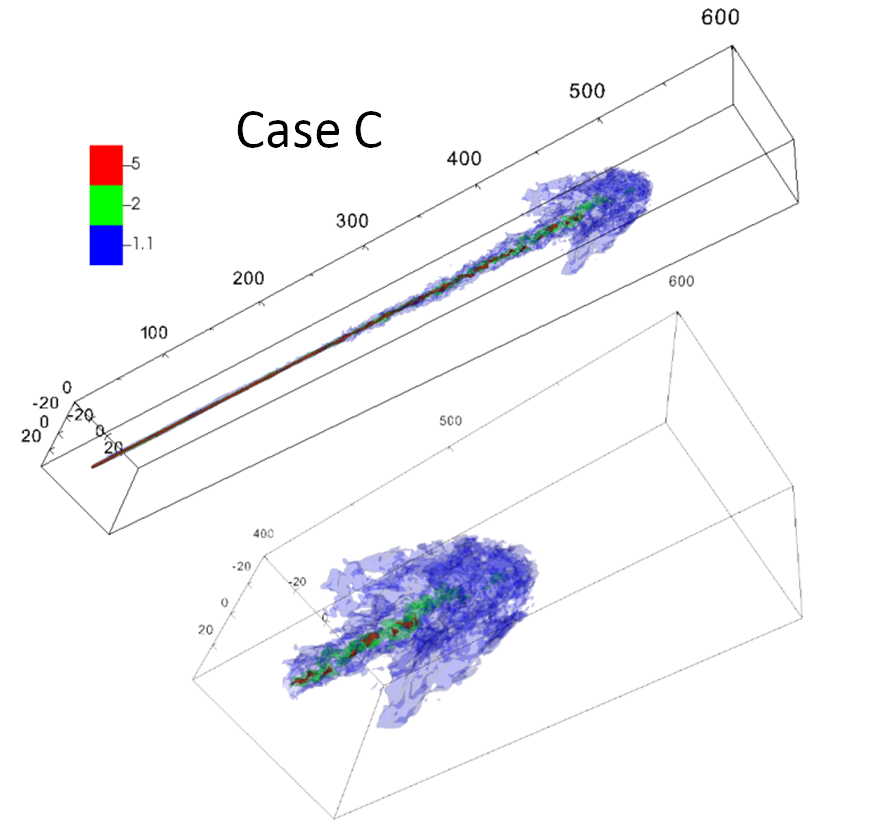}
    \includegraphics[width=\columnwidth]{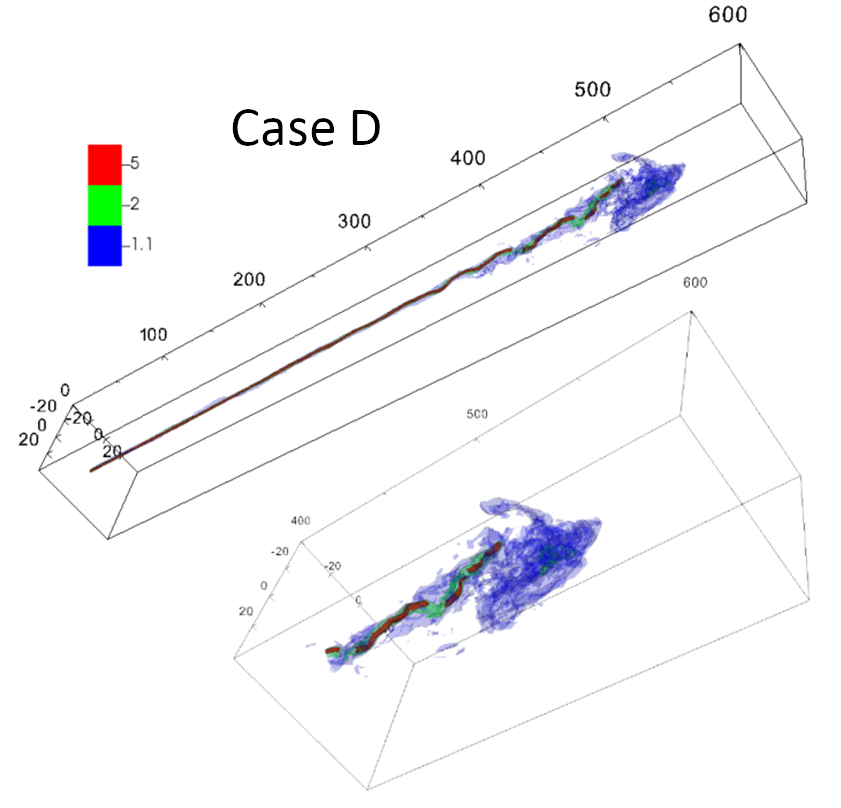}

    \caption{Three-dimensional isosurfaces of the Lorentz factor at the end of the simulations. The top-left panel represents case A at $t= 23000$, the top-right panel represents case B at $t = 12000$, the bottom-left panel represents case C at $t = 42000$ and, finally, the bottom-right panel represents case D at $t = 48000$. In each panel the top portions displays the entire jet while the bottom portions displays the last section of the jet between  $y=400$ and $y=600$. The three isosurfaces are for $\gamma = 1.1, 3, 5$. }
    \label{fig:gamma3D}
\end{figure*}

Our aim is to understand how the deceleration process occurs as a function of the density contrast and the jet magnetization. We then take two values for the density contrast, $\eta = 10^{-4}$ and $ 10^{-5}$, and two values for the jet magnetization, $\sigma_j = 10^{-2}$ and  $10^{-1}$, as shown in Table \ref{parset}. In principle the setup of our simulations is scale invariant; however, our interest is focused on the deceleration that mainly occurs in
the first kiloparsec. This motivates our assumption of constant
density of the external medium because the domain of our simulations can be assumed to be all contained within the galaxy
core. We can then adopt, as in Paper I, a jet injection radius of the order of one
parsec and the external density at injection of the order of one
particle per cubic centimeter \citep{balmaverde06}. In the following we will express all quantities in their
non-dimensional values as defined in the previous section. Conversion to physical units can be obtained with the above
assumptions; in particular, the unit of time is
\begin{equation}
    \tau \sim 3.26 \left( \frac{r_j}{1 \text{pc} }\right) \text{yrs} \,,
\end{equation}
and the jet power is

\begin{equation}\label{eq:jetpower}
    \begin{split}
    P_j &\sim h_0 \left(1 + 0.5 \sigma_j \right)\left( \frac{\gamma_j}{10}\right) \left( \frac{\eta}{10^{-4}}\right) \\ 
    & \left( \frac{r_j}{1 \text{pc}} \right)  \left( \frac{n_0}{1 \text{cm}^{-3}} \right) 1.22 \times 10^{43} \text{erg s}^{-1}\,,
    \end{split}
\end{equation}    
where $h_0$ is the relativistic specific enthalpy, defined in Eq. \ref{eq:enthalpy} and $n_0$ is the external number density. With the assumption of the jet Mach number $M_j = 2.2$, the value of $h_0$ is 1.6.   The jet power for the four cases is reported in Table \ref{tab:cases}. We choose the parameters so that the resulting powers are at about the dividing power between FR~I and FR~II.

We follow the jet propagation for cases A and B until the bow shock in front reaches $y=600$, in cases C and D we stop the simulations somewhat earlier because in these cases the cocoon edge reaches the lateral boundaries. In these cases the jet head at the end of the simulations reaches $y \sim 550$.

 In the following subsection we will discuss the general features of the jet dynamics, the entrainment and momentun transfer mechanisms and the resulting details of the jet deceleration.

\subsection{Jet dynamics}
In the two panels of Fig. \ref{fig:yhead} we show the jet head position as a function of time for all the cases: more precisely, the top panel shows the cases A and B, with density ratio $\eta_j = 10^{-4}$, and the bottom panel shows the cases C and D, with density ratio $\eta_j = 10^{-5}$. The dashed black line in each panel shows the estimated head position obtained by using  the theoretical value of the jet head velocity, in the relativistic case,  given by \citet{Marti97} \citep[see also][]{Bromberg11}. This value has been obtained, by equating the (relativistic) momentum flux of the jet and the ambient medium, as
\begin{equation}
    v_h = \frac{\gamma_j \sqrt{\eta_R}}{1+ \gamma_j  \sqrt{\eta_R}}\,, \qquad\qquad
    \eta_r = \frac{h_0}{h_a} \eta\,,
\end{equation}\label{eq:vh}  
where $h_0$ and $h_a$ are, respectively, the specific relativistic enthalpies of the jet and of the ambient. In the top panel, we see that case A (blue curve, lower magnetization $\sigma_j = 0.01$) follows closely the theoretical estimate up to $y_h \sim 300$ ($t \sim 2500$).  Actually the jet-head velocity reaches values that are slightly higher than the theoretical estimate and this could be interpreted as the result of jet-head wobbling as in \citet{aloy99} and \citet{perucho19}. After this initial phase,  we observe a strong deceleration. In Case B (orange curve, higher magnetization $\sigma_j = 0.1$), at the beginning, the jet head appears to advance more slowly than in case A, however the deceleration is much less efficient and the jet head reaches the end of the domain in about half the time with respect to case A ($t\sim 12000$ for case B vs $t\sim 23000$ for case B). In the bottom panel, cases C and D propagate at a much smaller velocity, as it is expected from the lower density ratio. Case C, with lower magnetization ($\sigma_j = 0.01$, blue curve), at the beginning, propagates  faster than case D (orange curve, higher magnetization $\sigma_j = 0.1$), but, at later times, case C shows a stronger deceleration and the two cases reach the end of the domain at similar times ($t \sim 45000$).

\begin{figure*}
    \centering
    \includegraphics[width=\columnwidth]{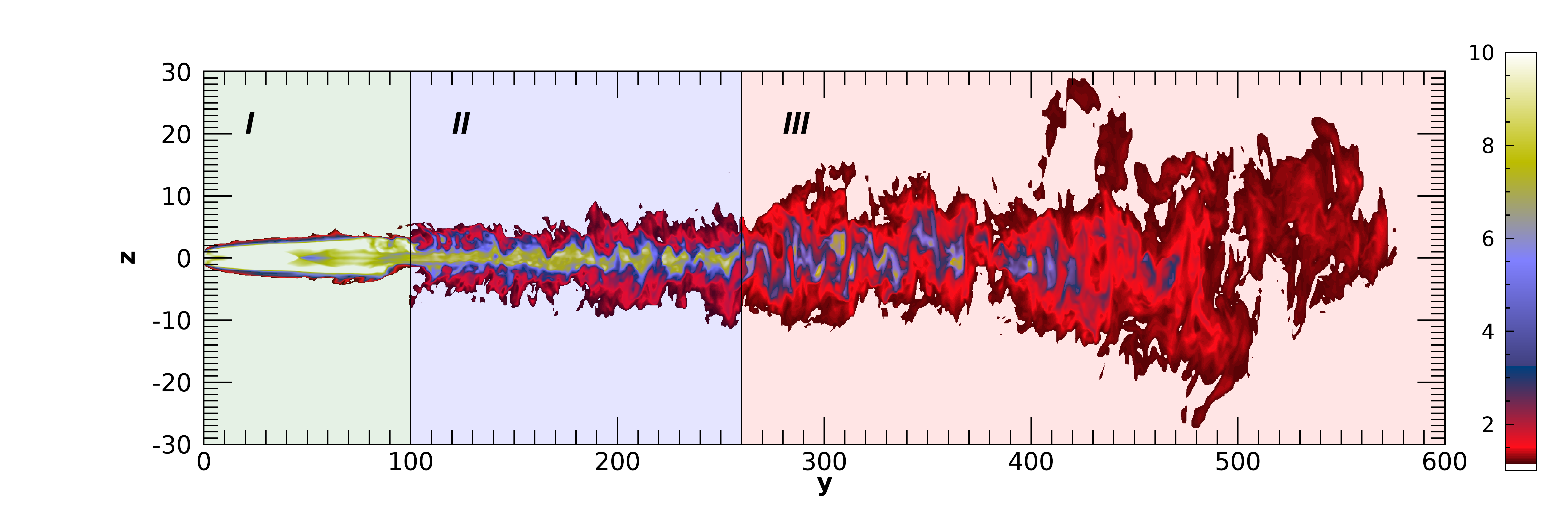}
    \includegraphics[width=\columnwidth]{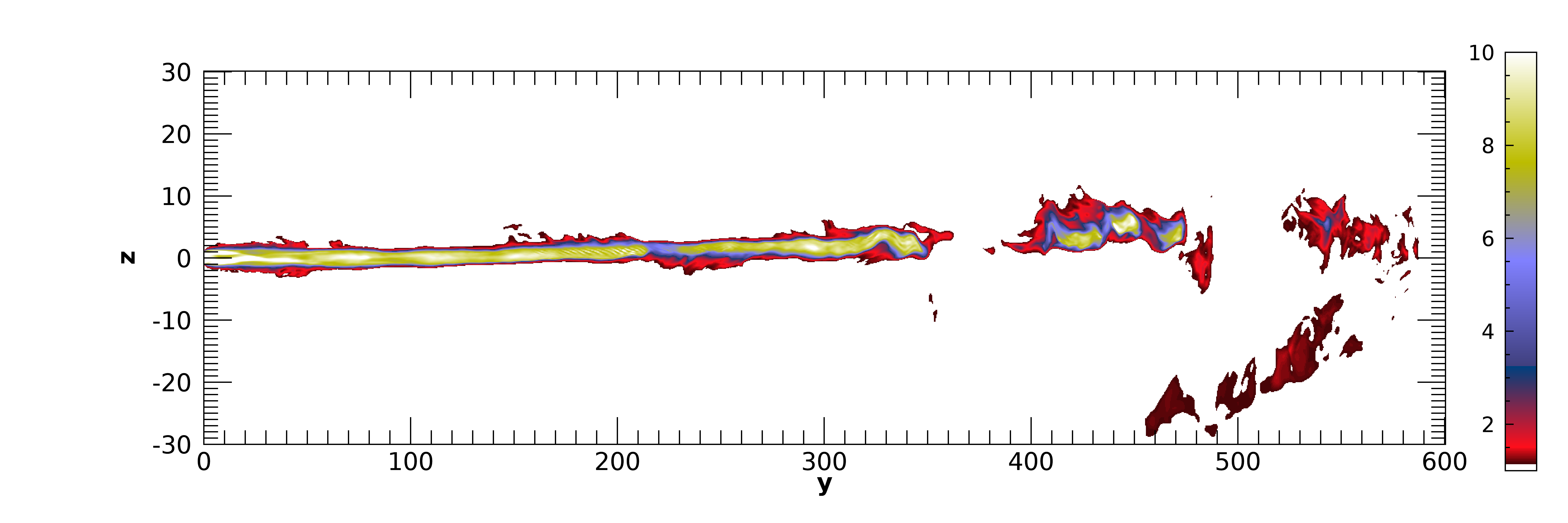} \\
    \includegraphics[width=\columnwidth]{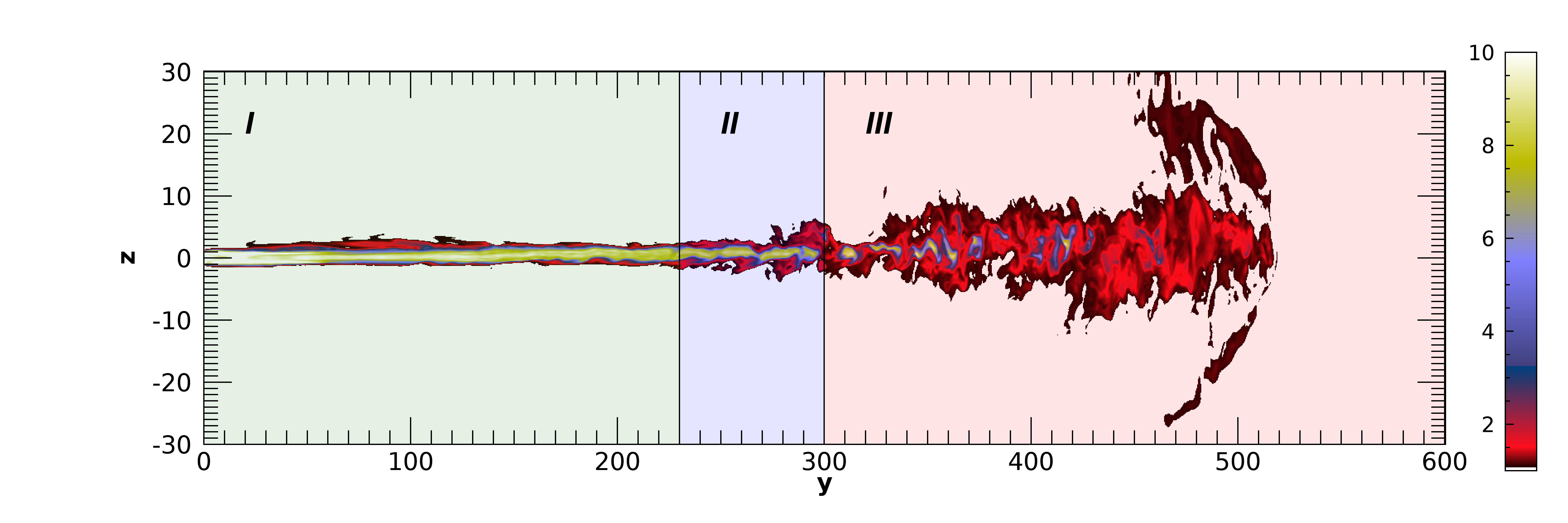}
    \includegraphics[width=\columnwidth]{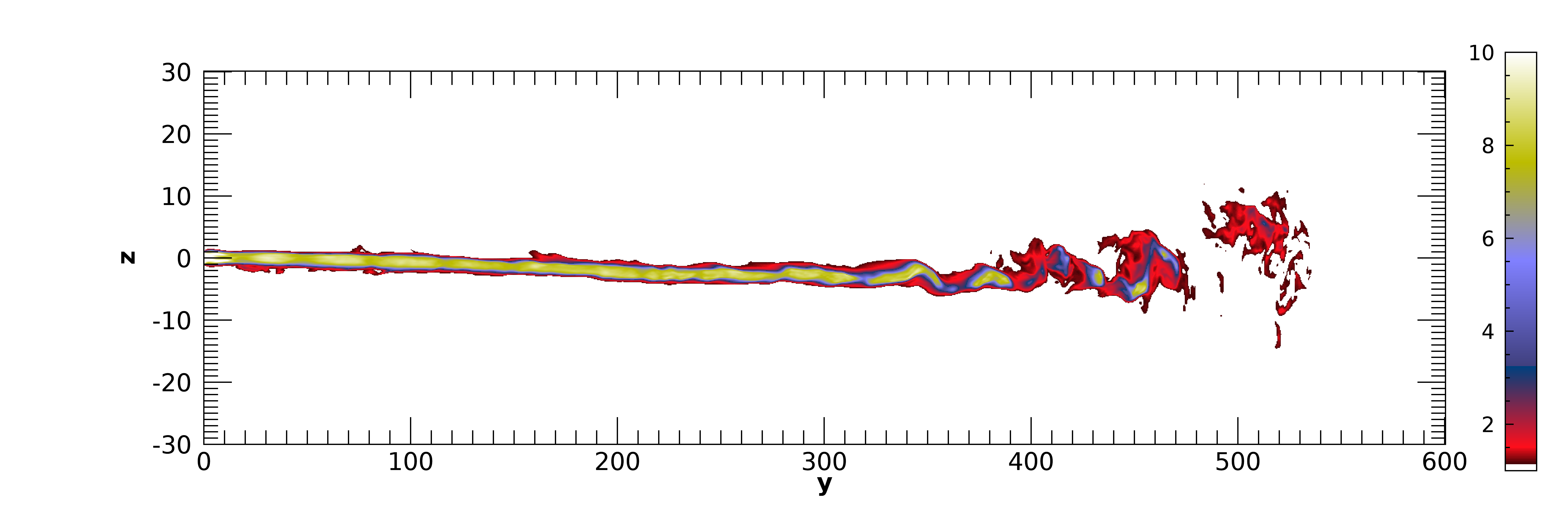}

    \caption{Longitudinal cut (in the $y-z$ plane) of the distribution of the Lorentz factor at the final times of the simulations. The four panels refer to the four different cases (Case A top-left, Case B top-right, Case C bottom-left,  Case D bottom-right).  The times are the same as in Fig. \ref{fig:gamma3D}. The three coloured bands in the left panels (Cases A and C) highlight the three phases described in the text.}
    \label{fig:gcut}
\end{figure*}

\begin{figure}
    \centering
    \includegraphics[width=\columnwidth]{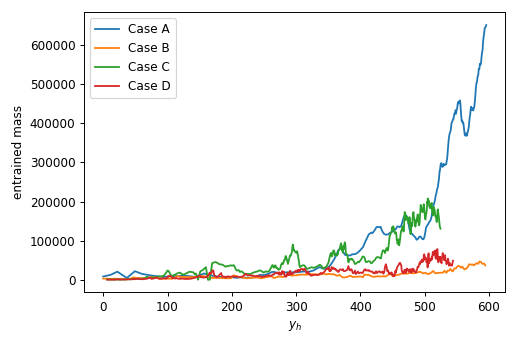}

    \caption{Plot of the entrained mass as a function of the jet head position for all the cases. The entrained mass is defined as the mass of the external material moving at $v_y > 0.1$}
    \label{fig:entr}
\end{figure}

\begin{figure*}
    \centering
    \includegraphics[width=\columnwidth]{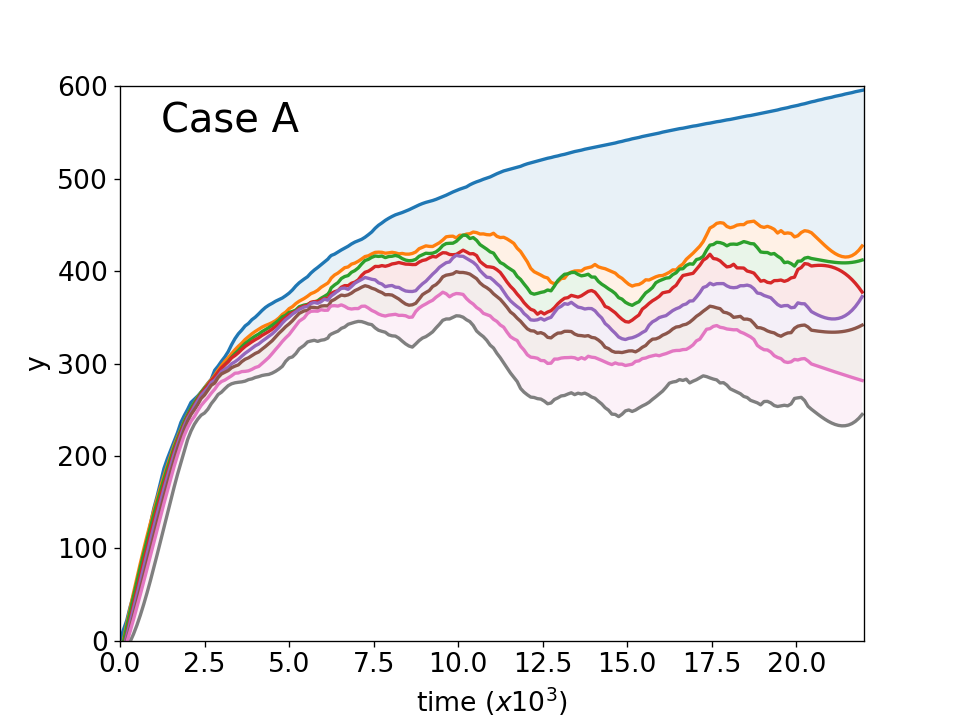}
    \includegraphics[width=\columnwidth]{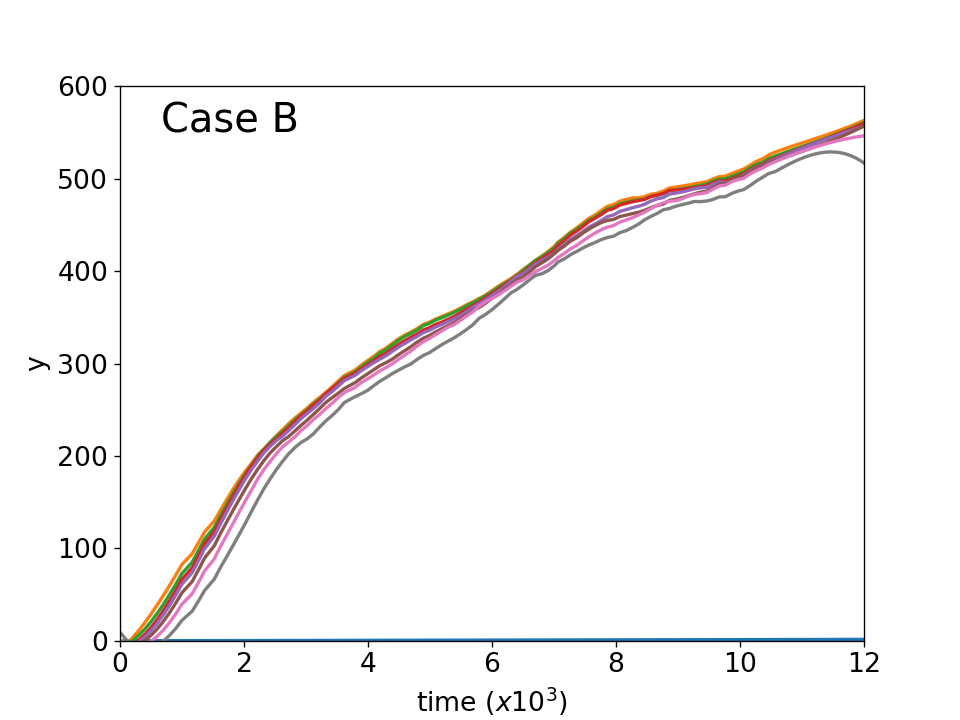} \\
    \includegraphics[width=\columnwidth]{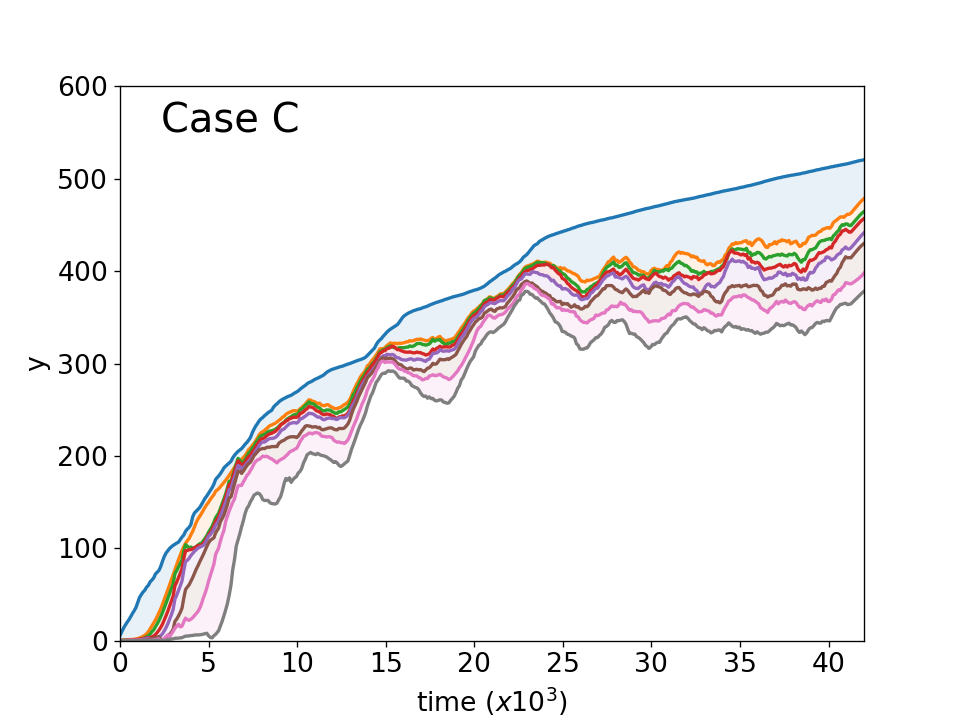}
    \includegraphics[width=\columnwidth]{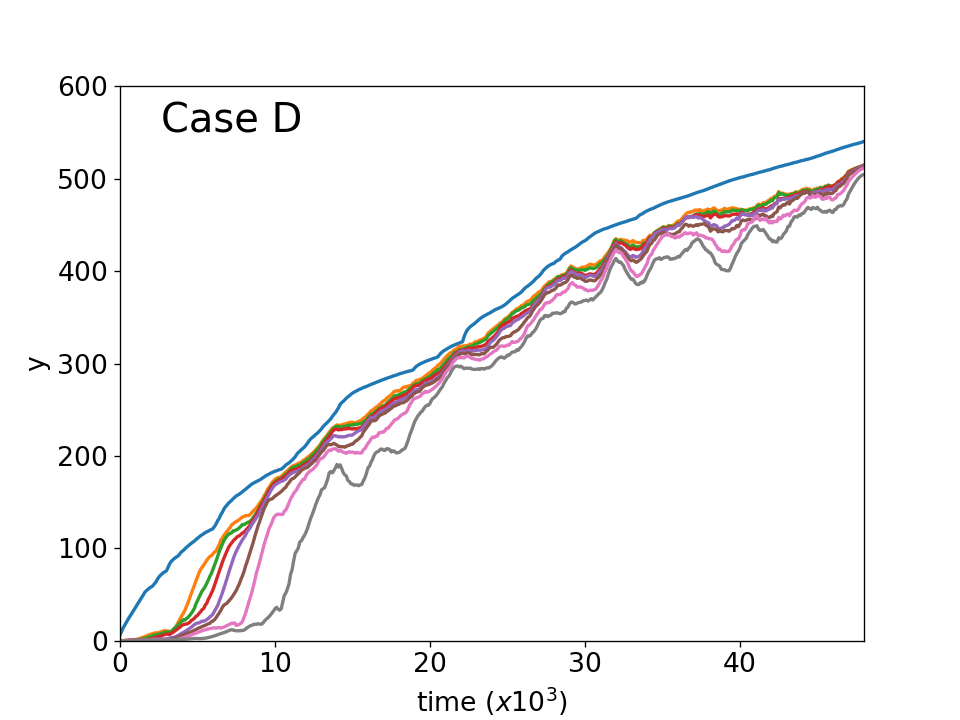}

    \caption{Maximum distance from the jet origin at which a given fraction of momentum flux is carried by material moving at $\gamma \geq 5$ as a function of time. The four panels refer to the four different cases (Case A top-left; Case B top-right; Case C bottom-left; Case D bottom-right). In each panel the top curve shows the jet head position and the other curves represent the fractions  from 10\% to 70\%. For cases A and C we highlight the areas between subsequent curves with different colors.  }
    \label{fig:momfrac}
\end{figure*}

From Fig. \ref{fig:gamma3D} we can get an impression of the final structure of the jets. In the figure, we show, for all the cases, the three-dimensional distribution of the Lorentz factor, when the jet reached its maximum length. The four snapshots corresponds to four different times, namely $t \sim 23000$ for case A, $t \sim 12000$ for case B, $t \sim 42000$ for case C and $t \sim 48000$ for case D. The differences in times are due to the different jet head propagation velocities, as we saw above. In the figures, we display three-dimensional views of the isocontours for three values of the Lorentz gamma factor ($\gamma = 1.1, 3, 5$). Each panel shows both the entire jet and a zoom on the last 200 length units. The represented domain, in the transverse direction, extends from $-30$ to $30$, in both directions, for all cases except case A, that extends from $-40$ to $40$, since, in this case we observe a larger expansion of the jet. In cases A and C, which are the less magnetized, we have a behavior similar to the hydrodynamic cases discussed in Paper 2. We observe a first region, in which the jet seems to propagate straight, then (for $y \gtrsim 100$ in case A and $y \gtrsim 250 $ in case C) we observe the growth of perturbations, with the appearance of jet wiggling, and eventually the jet breaks into small high velocity fragments surrounded by slow moving material.  Notice that we do not add any perturbation at the jet inlet and the growth of perturbations is solely due to the jet-backflow interaction. In case A, the zoom on the last part of the jet shows that material moving at $\gamma > 5$ is practically absent. The behavior of the more magnetized cases (B and D in the right panels) is quite different. The high velocity part of the jet remains more coherent, with some wiggling, but at a larger scale with respect to the other two cases. The jet fragmentation occurs but the high velocity fragments have a larger size and are present up to the jet head.

\begin{figure*}[!ht]
    \centering
    \includegraphics[width=\columnwidth]{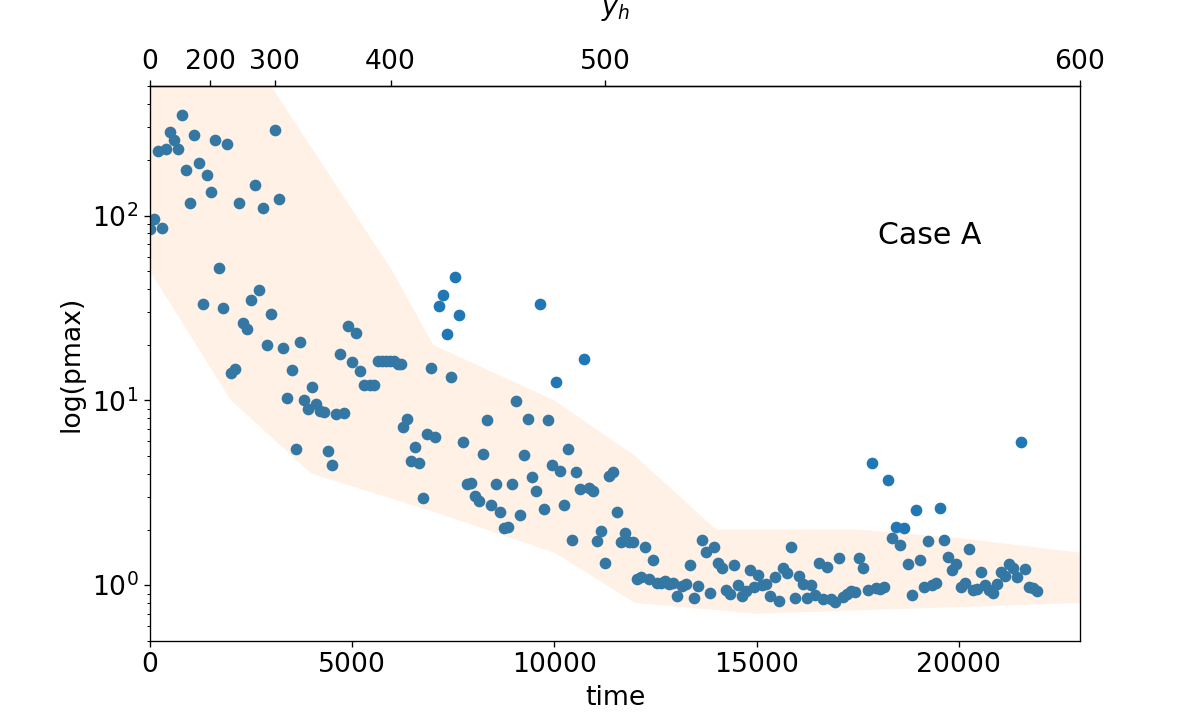}
    \includegraphics[width=\columnwidth]{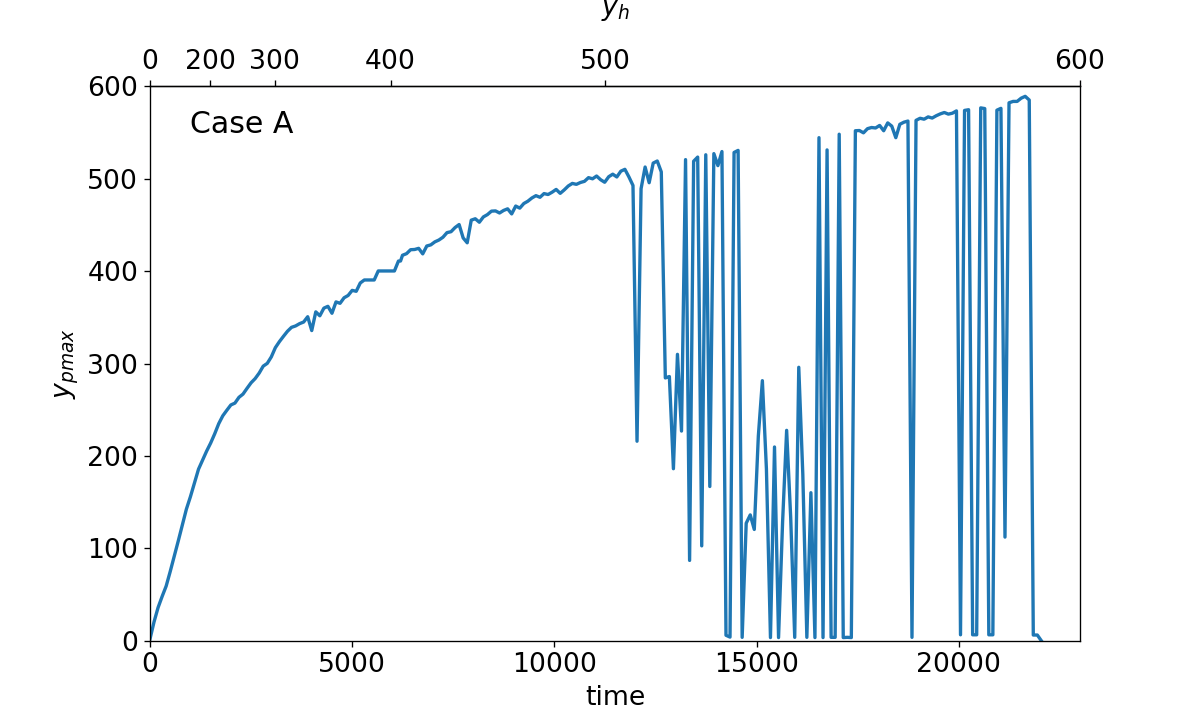} \\
    \includegraphics[width=\columnwidth]{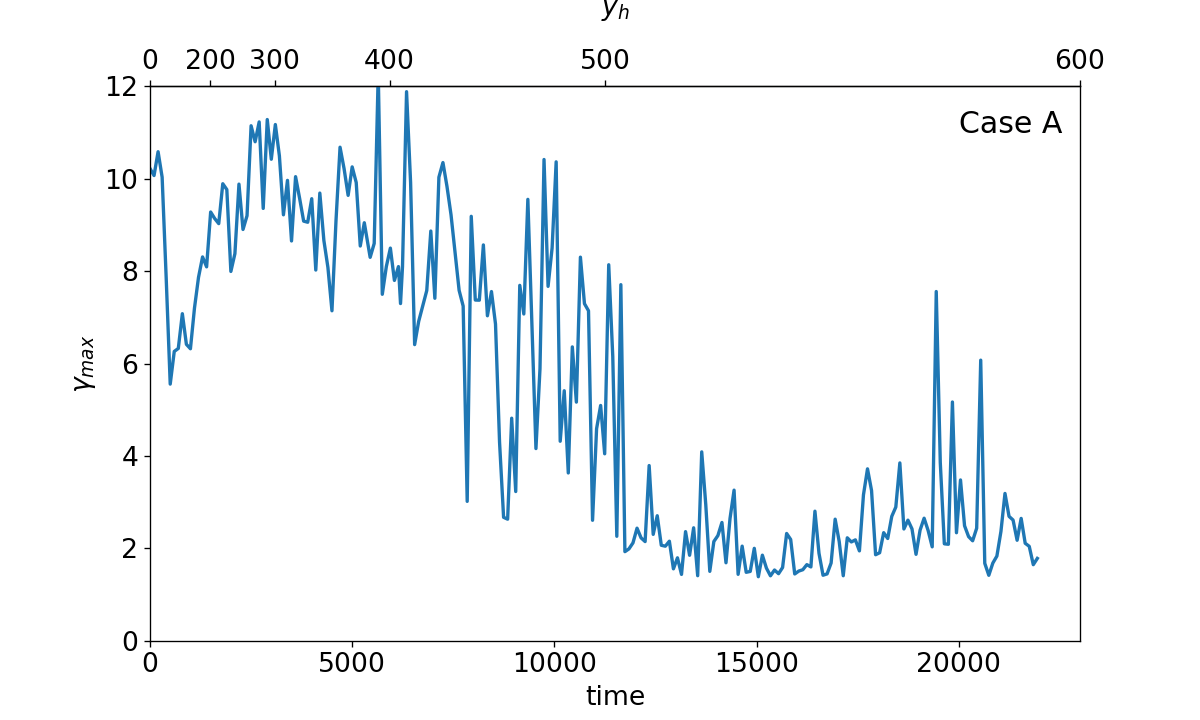}
    \includegraphics[width=\columnwidth]{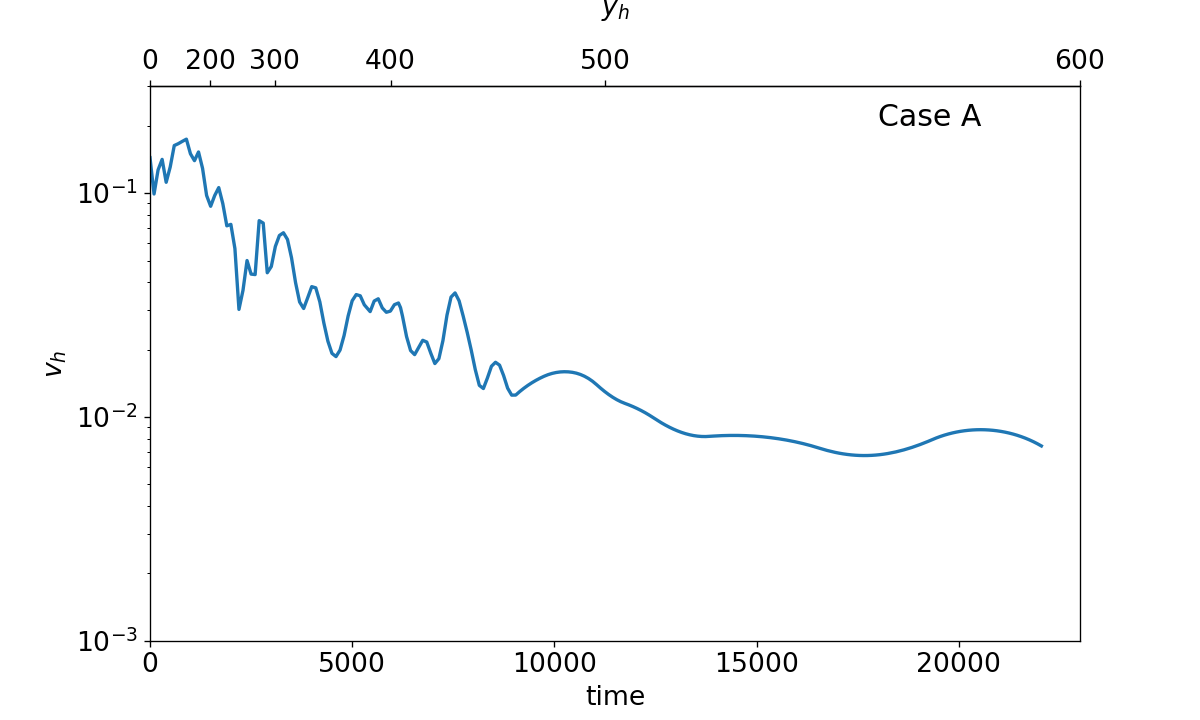}

    \caption{Case A - Top-left panel: maximum pressure found in the computational domain as a function of time (bottom axis) and the position reached by the jet head (top axis). The symbols mark the values for each time, while the colored band identifies the range of variation. Top-right panel: plot of the y coordinate of maximum pressure, plotted in the top-left panel, as a function of time (bottom axis) and the position reached by the jet head (top axis). Bottom-left panel: plot of the maximum Lorentz factor near the jet head as a function of time (bottom axis) and the position reached by the jet head (top axis). More precisely the maximum value of $\gamma$ is in the region $y_h -50 < y < y_h$, where $y_h$ is the jet head position. Bottom-right panel: plot of the jet head velocity as a function of time (bottom axis) and the position reached by the jet head (top axis).  }
    \label{fig:compositeA}
\end{figure*}

\begin{figure*}
    \centering
    \includegraphics[width=\columnwidth]{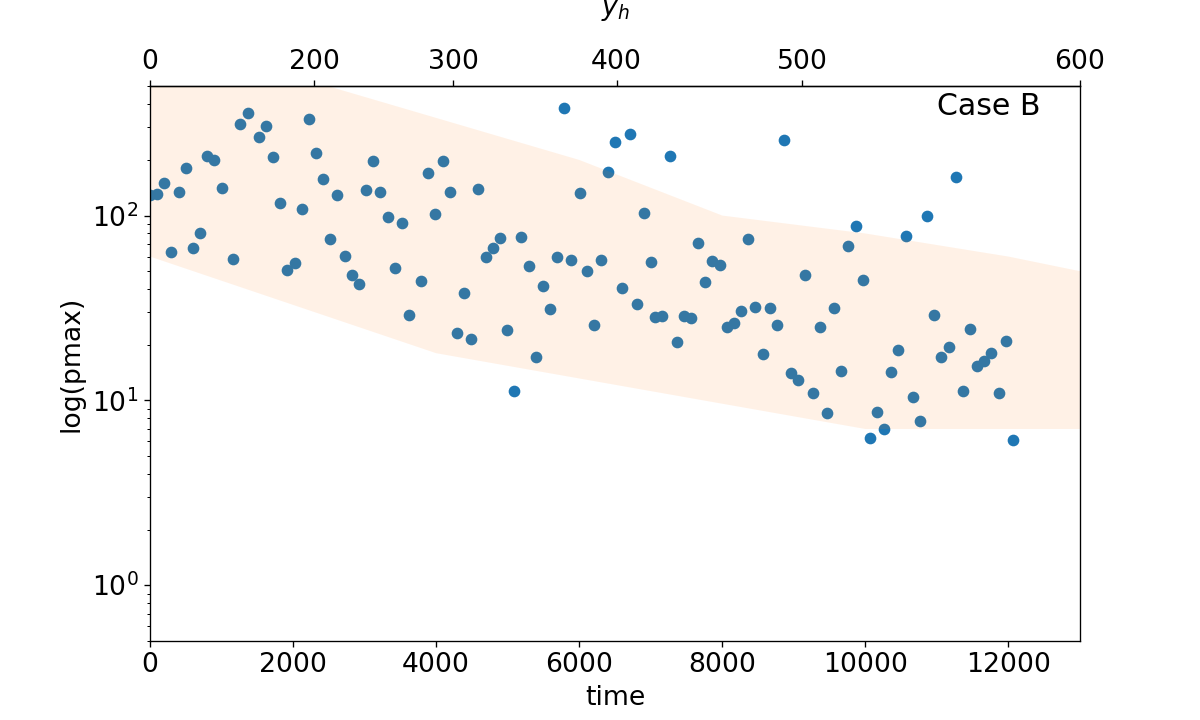}
    \includegraphics[width=\columnwidth]{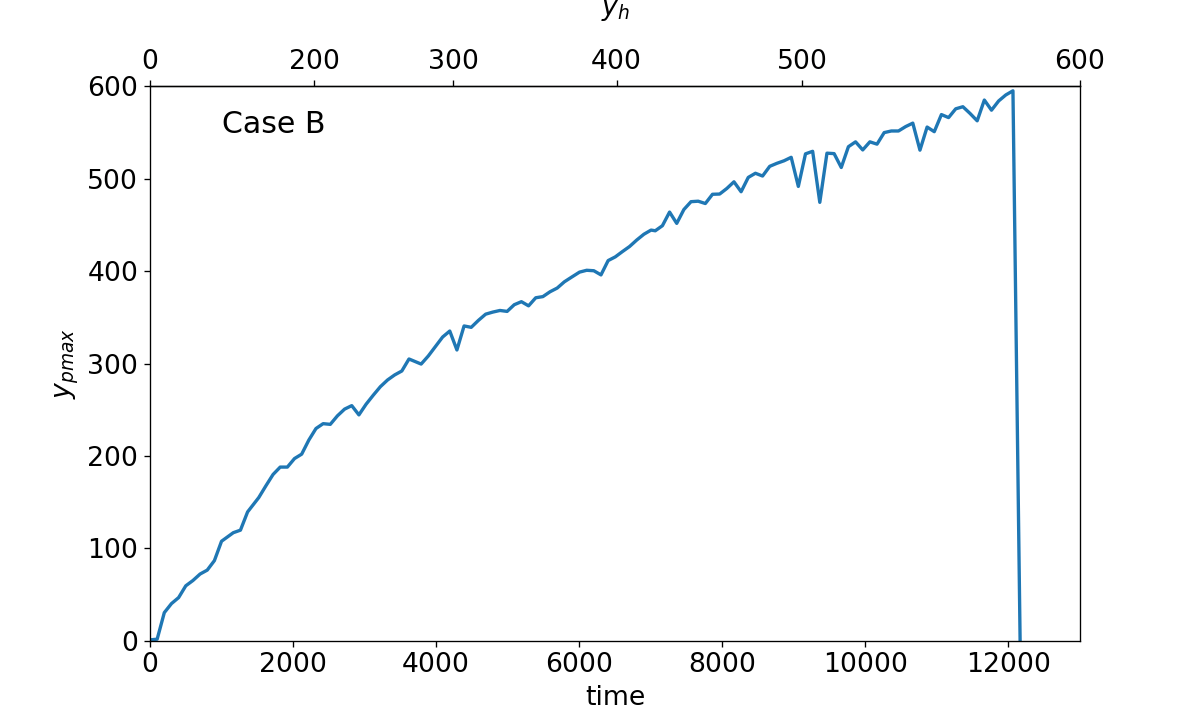} \\
    \includegraphics[width=\columnwidth]{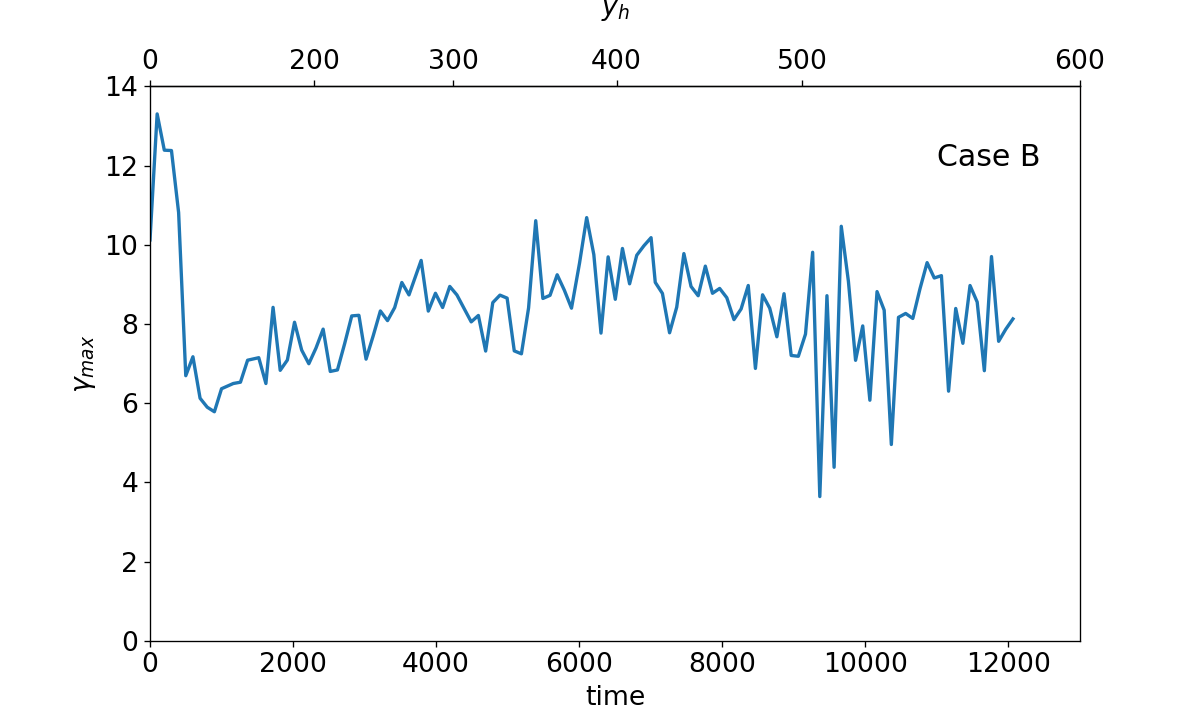}
    \includegraphics[width=\columnwidth]{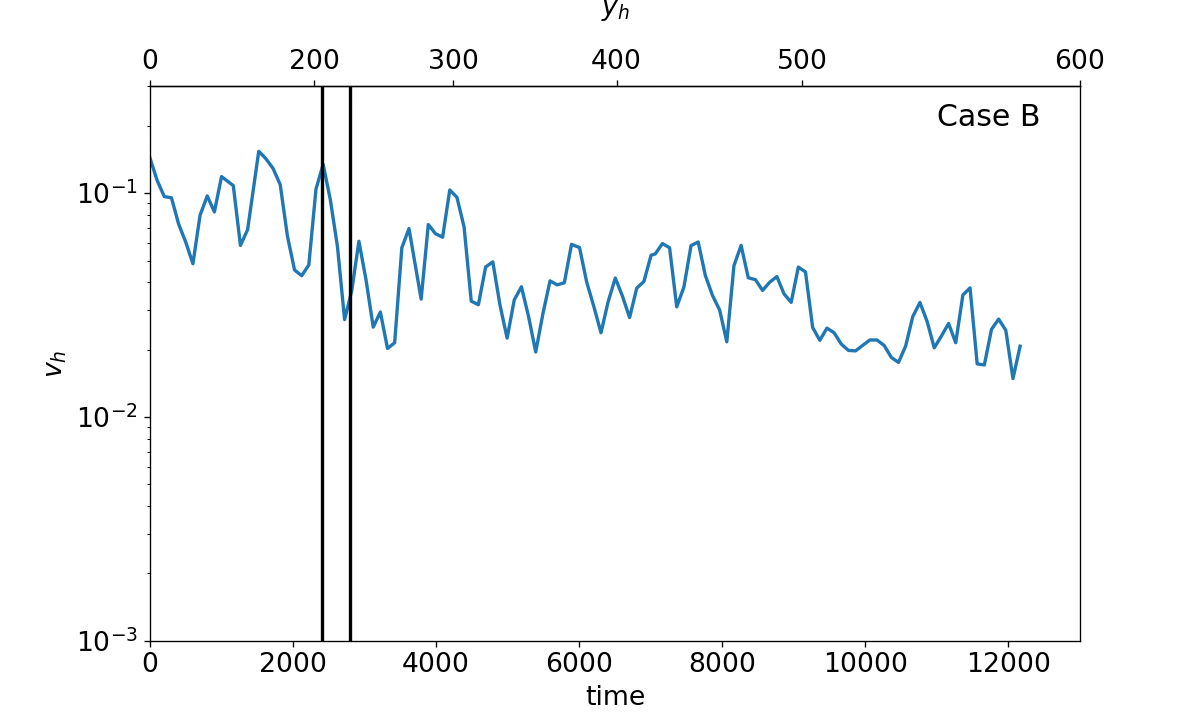}

    \caption{Same as Fig. \ref{fig:compositeB}, but for Case B. The two vertical black lines in the bottom-right panel indicate one of  time interval in which the jet slows down because of the bending of the jet head as shown in Fig. \ref{fig:kink}  }
    \label{fig:compositeB}
\end{figure*}

\begin{figure*}
    \centering
    \includegraphics[width=\columnwidth]{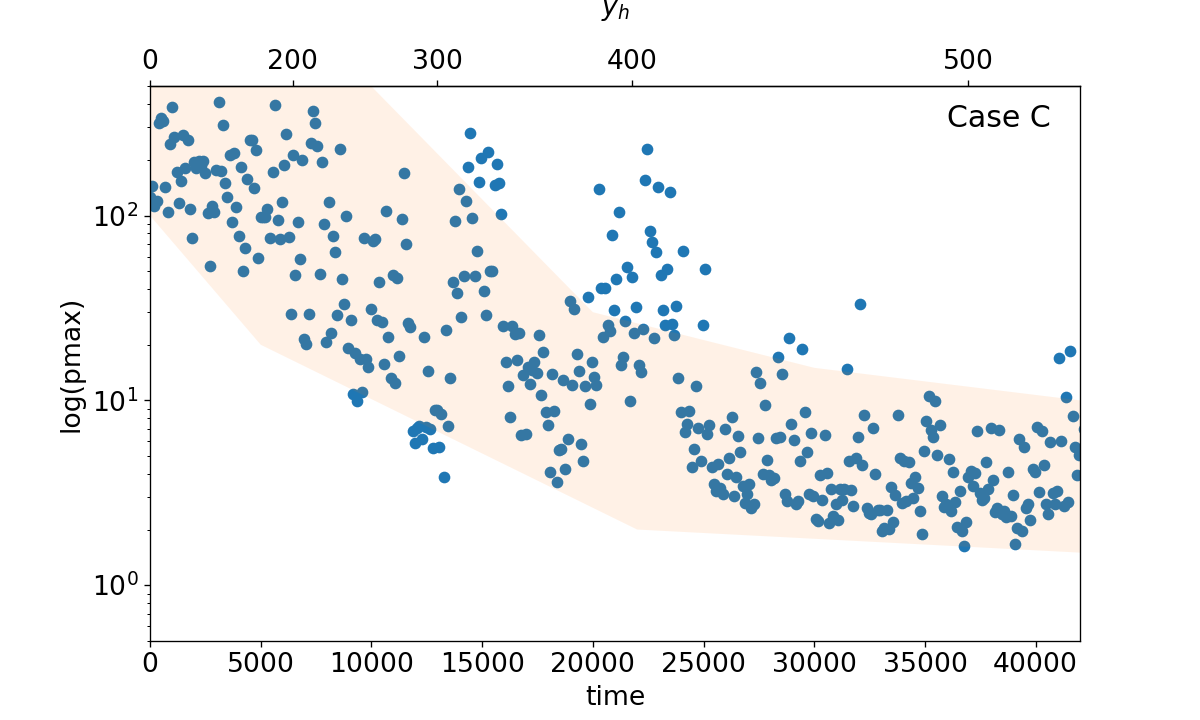}
    \includegraphics[width=\columnwidth]{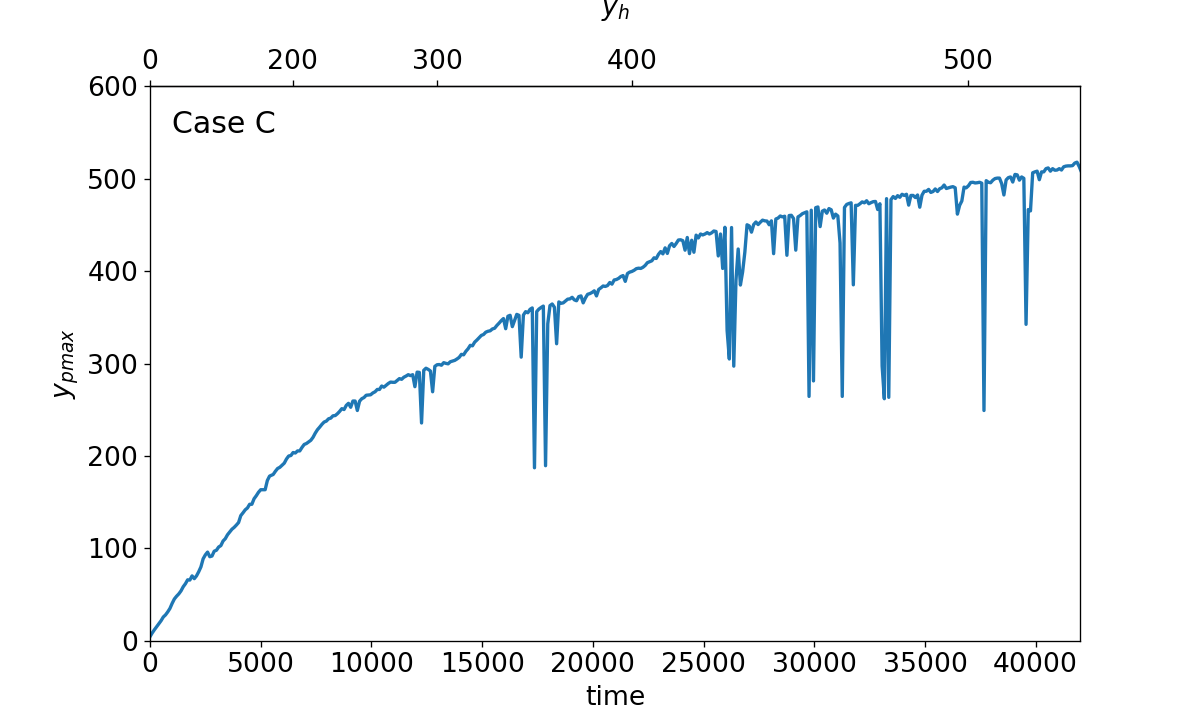} \\
    \includegraphics[width=\columnwidth]{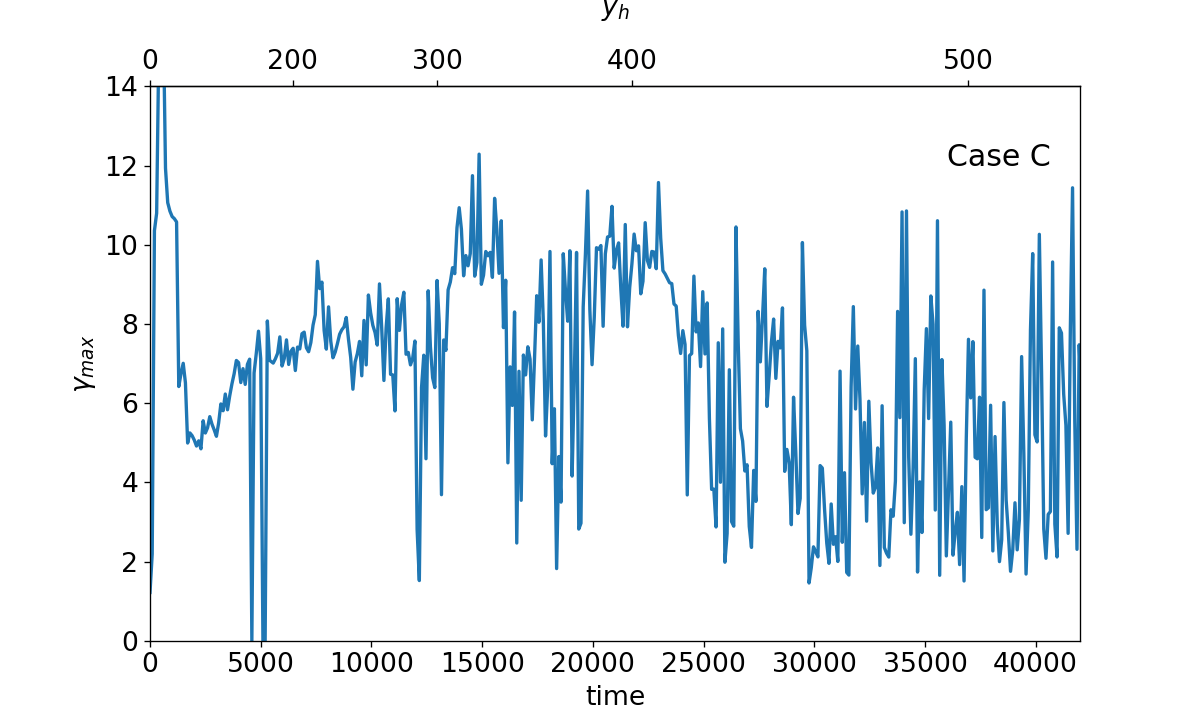}
    \includegraphics[width=\columnwidth]{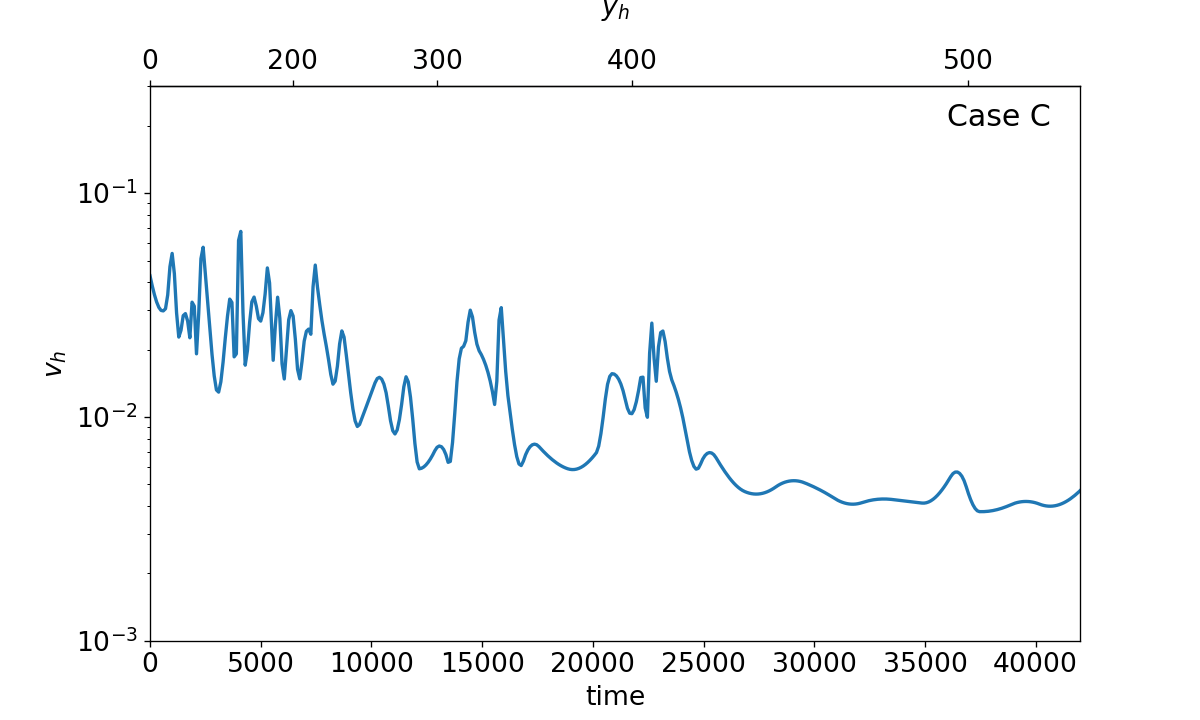}

    \caption{Same as Fig. \ref{fig:compositeB}, but for Case C. }
    \label{fig:compositeC}
\end{figure*}

The behavior described above can be better appreciated in Fig. \ref{fig:gcut}, where we display a longitudinal cut (in the $y-z$ plane) of the distribution of the Lorentz factor at the final times of the simulations. In Paper II we described three regions along the jet characterized by a different jet behavior. In the first region the jet propagates almost undisturbed and we observe the presence of several recollimation shocks, in the second region perturbations grow until, in the last region, the jet is
broken into a few fast-moving blobs surrounded by a wide low-velocity envelope. In the present simulations, these regions are clearly visible  for cases A and C: in case A, the unperturbed region is quite short, up to $y \sim 100$, and characterized by  strong recollimation shocks, the second region spans the jet length from $y \sim 100$ to $y \sim 250$ and for $y \gtrsim 250$ the jet is fragmented; in case C, the first region ends at $y \sim 250$ and the recollimation shocks are quite weak, the second region ends at $y \sim 350$ and the fragmentation region is observed for $y \gtrsim 350$. A comparison with the results of Paper II  (see Case C of Paper II with a density ratio of $10^{-4}$) shows that in case A we have a much faster growth of perturbations and an earlier jet fragmentation, while case C behaves similarly to the hydro case. In addition we can also notice that, in case A, the entrainment region is much wider with respect to case C (and the hydro cases). Cases B and D, which are more magnetized, show a quite different behavior, the jet propagates quite straight for larger distances, up to $y \gtrsim 300$ and then appears to break more suddenly, with the formation of larger fragments and a much lower entrainment. 

\subsection{Jet entrainment and momentum transfer}

The results presented above show that, in cases A and C, the jet progressively
decelerates and the quantity of material moving at high $\gamma$ decreases, while an increasing amount of material around the jet moves at $\gamma \leq 2$. Now we can analyze more quantitatively how the entrainment and  deceleration processes occur.
In Fig. \ref{fig:entr} we show the behavior of the entrained mass as a function of the jet head position, more precisely we define the entrained mass as the total mass of the external material moving with $v_y > 0.1$. The four curves in the figure refer to the four different cases. Case A, as expected, is the one with the highest value of the entrained mass. Most of the entrainment occurs for $400 < y_h < 600$, that corresponds to $7000 < t < 24000$, i.e. to the largest fraction of the duration of the jet evolution. For Case C we had to stop the simulation somewhat earlier with respect to case A because the lateral bottom part of the cocoon exits the computational grid. Up to this point, however, case C behaves in a way similar to Case A.  For the more magnetized cases B and D, instead,  the entrained mass is much lower,  less than  $10\%$ of that of case A.  

\begin{figure*}
    \centering
    \includegraphics[width=\columnwidth]{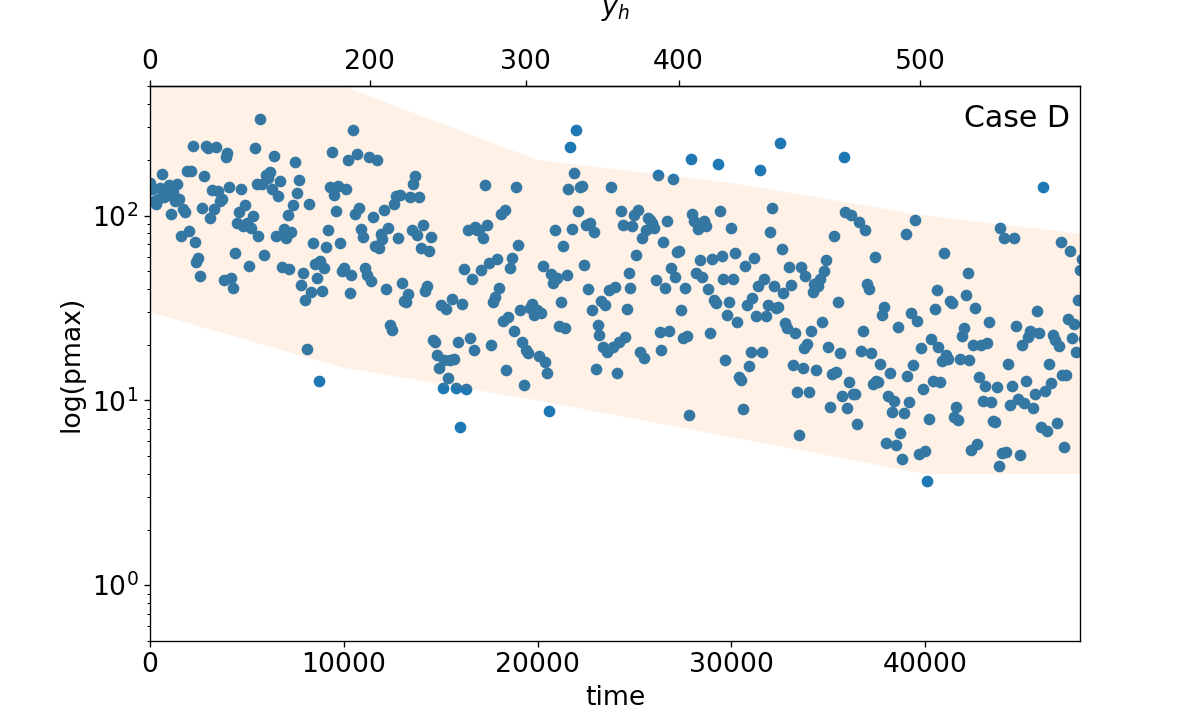}
    \includegraphics[width=\columnwidth]{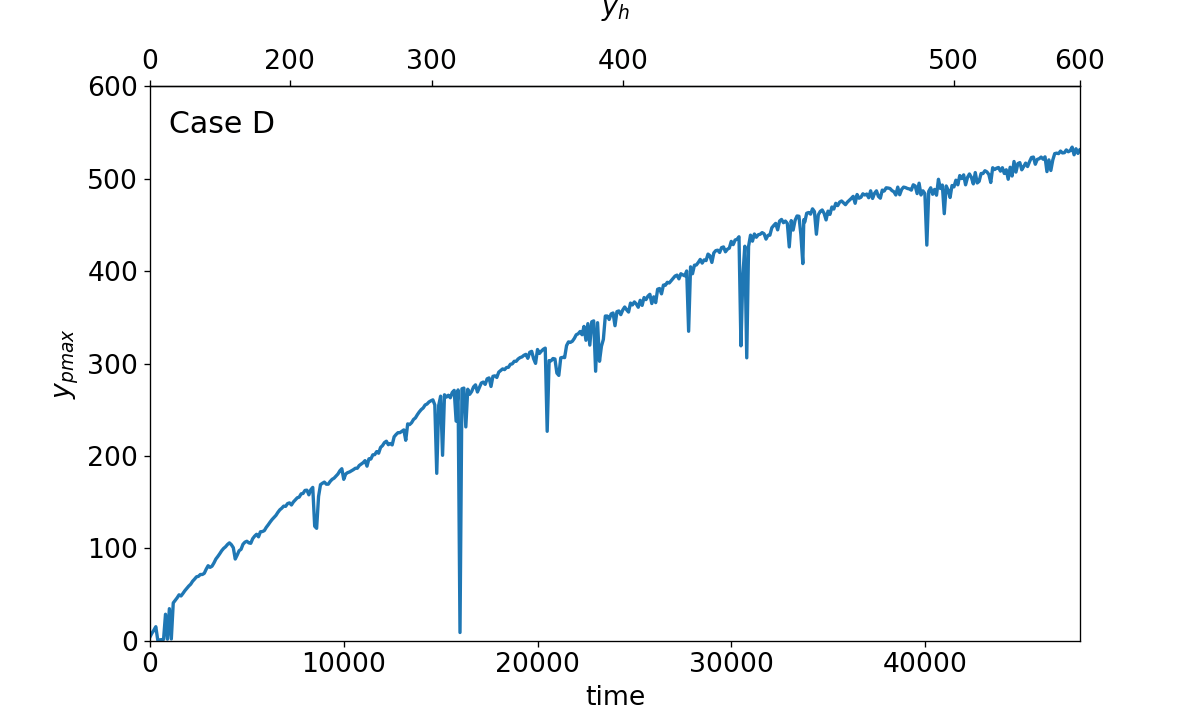} \\
    \includegraphics[width=\columnwidth]{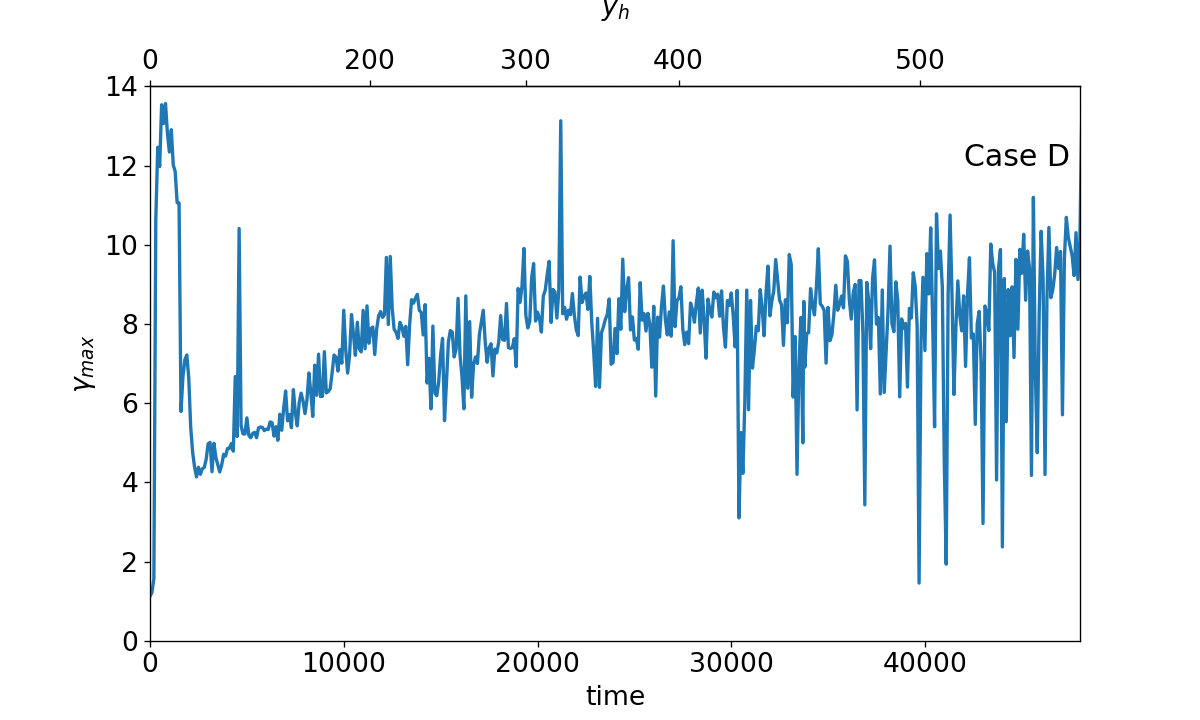}
    \includegraphics[width=\columnwidth]{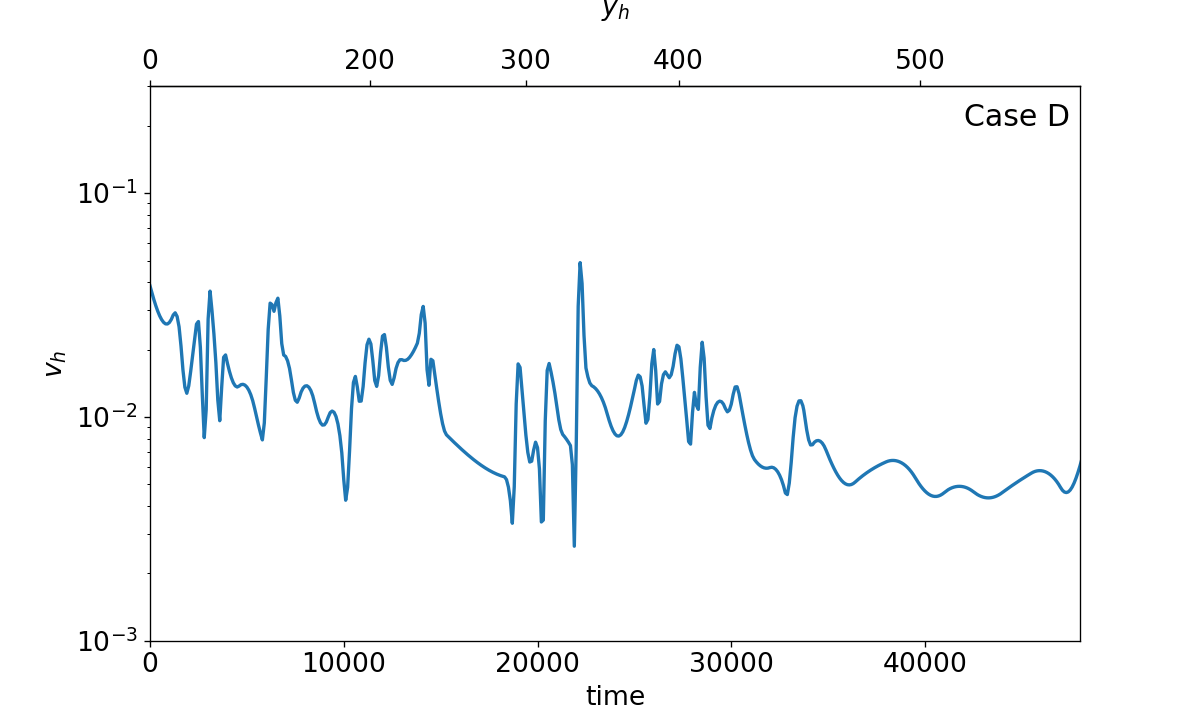}

    \caption{Same as Fig. \ref{fig:compositeB}, but for Case D.  }
    \label{fig:compositeD}
\end{figure*}

\begin{figure}
    \centering
    \includegraphics[width=4.8cm]{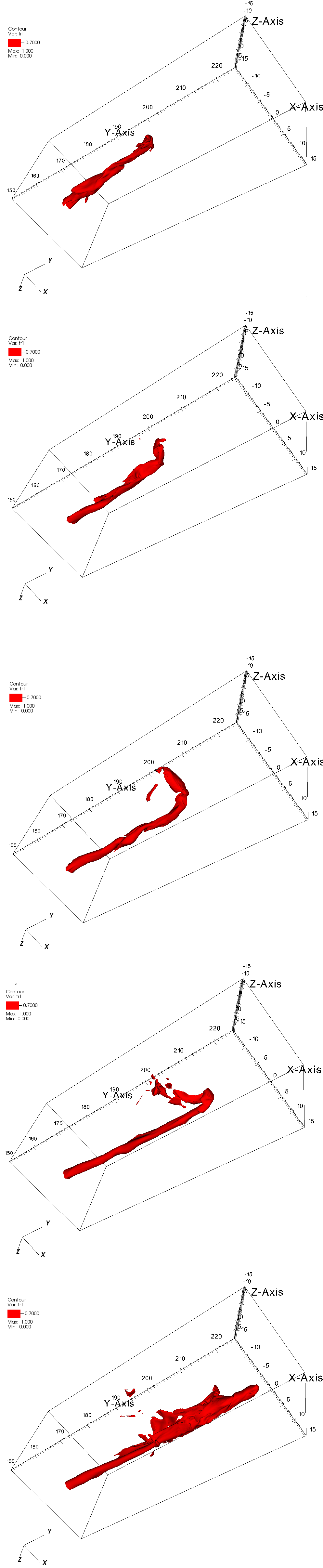}

    \caption{The five panels show  three-dimensional iso-contours of the tracer (for a tracer value of 0.7), for case B,  at five different times in the interval marked by the black vertical lines in the top-right panel of Fig. \ref{fig:compositeB}.  }
    \label{fig:kink}
\end{figure}

\begin{figure*}
    \centering
    \includegraphics[width=\columnwidth]{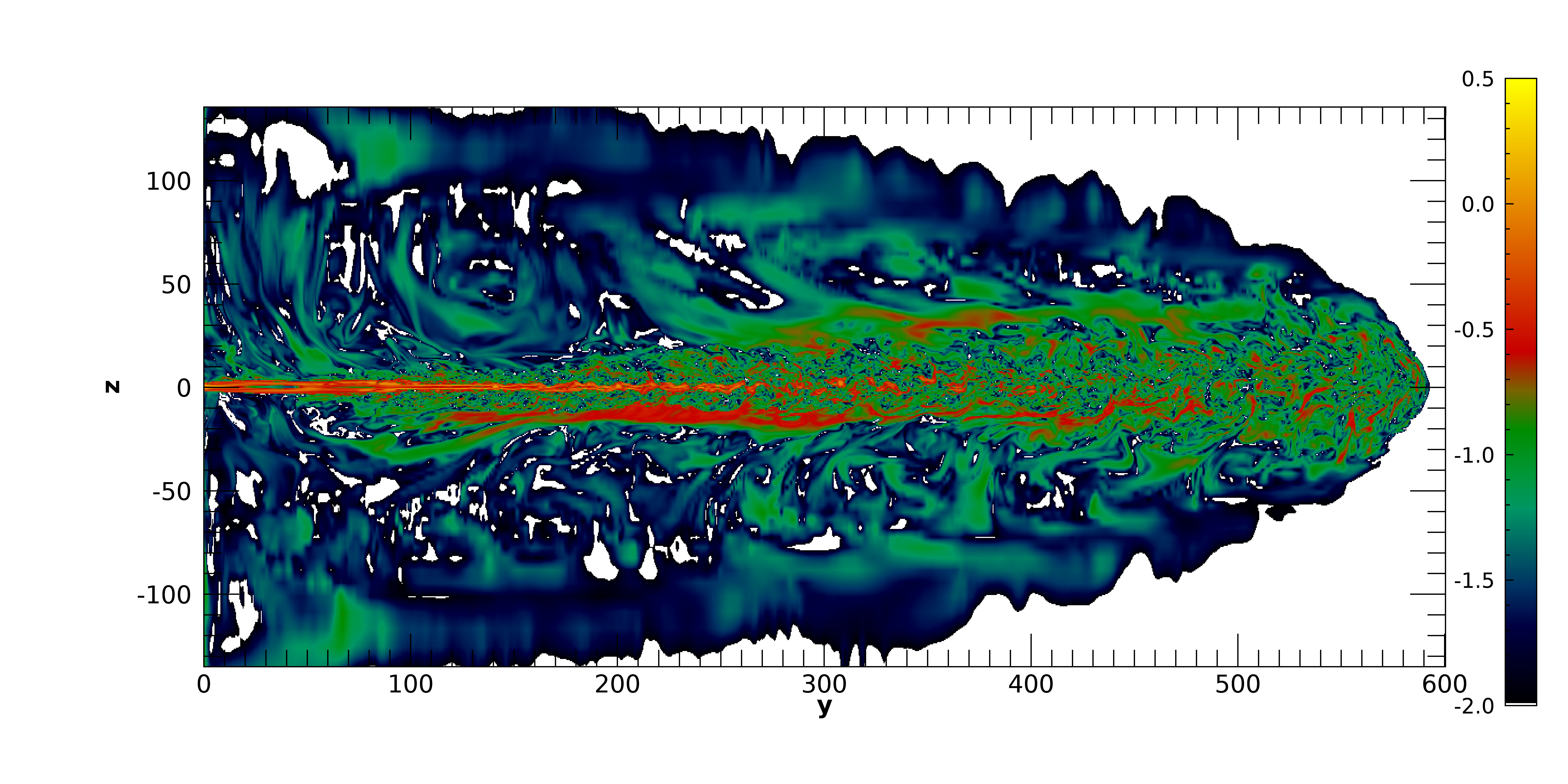}
    \includegraphics[width=\columnwidth]{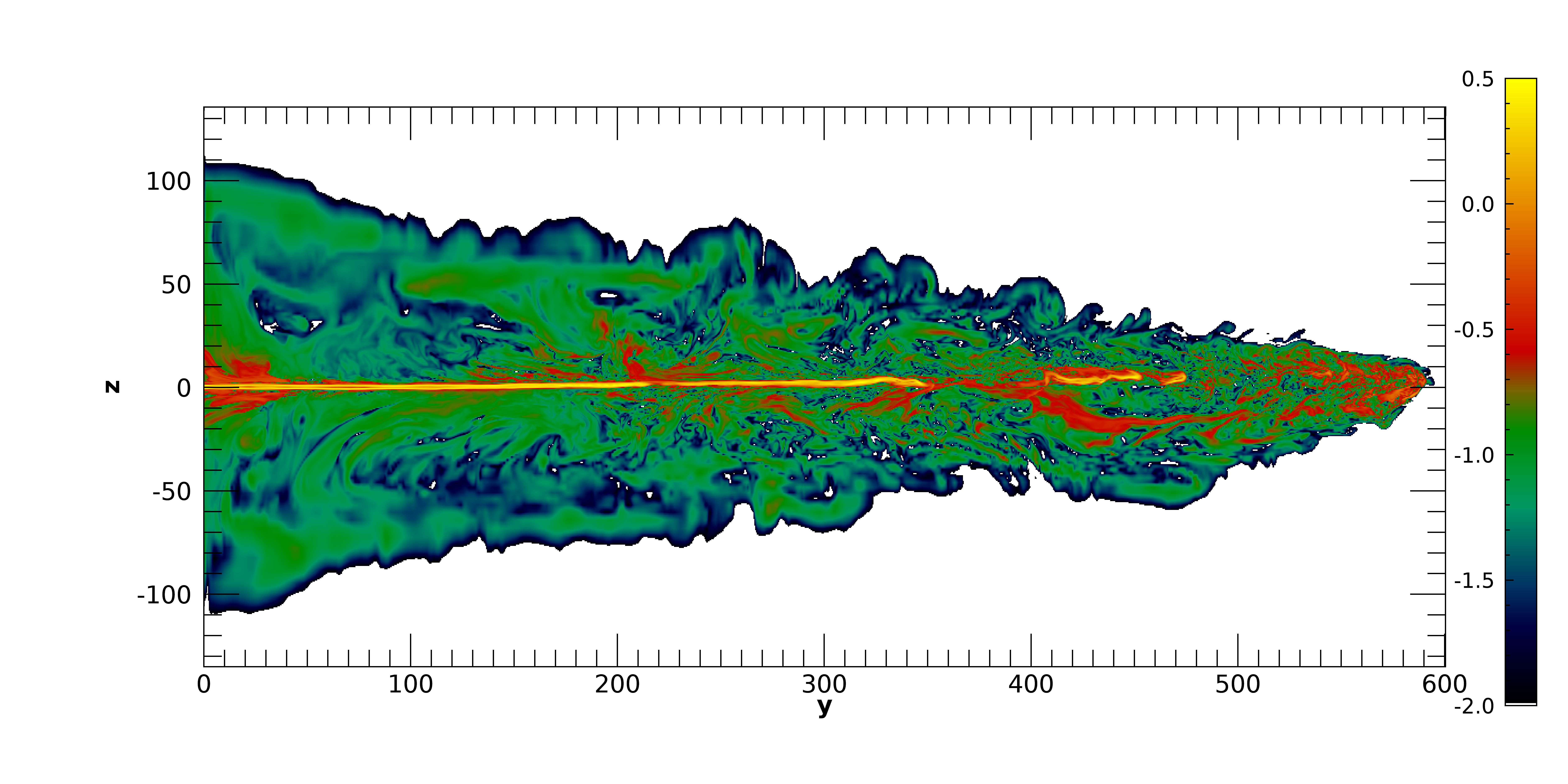} \\
    \includegraphics[width=\columnwidth]{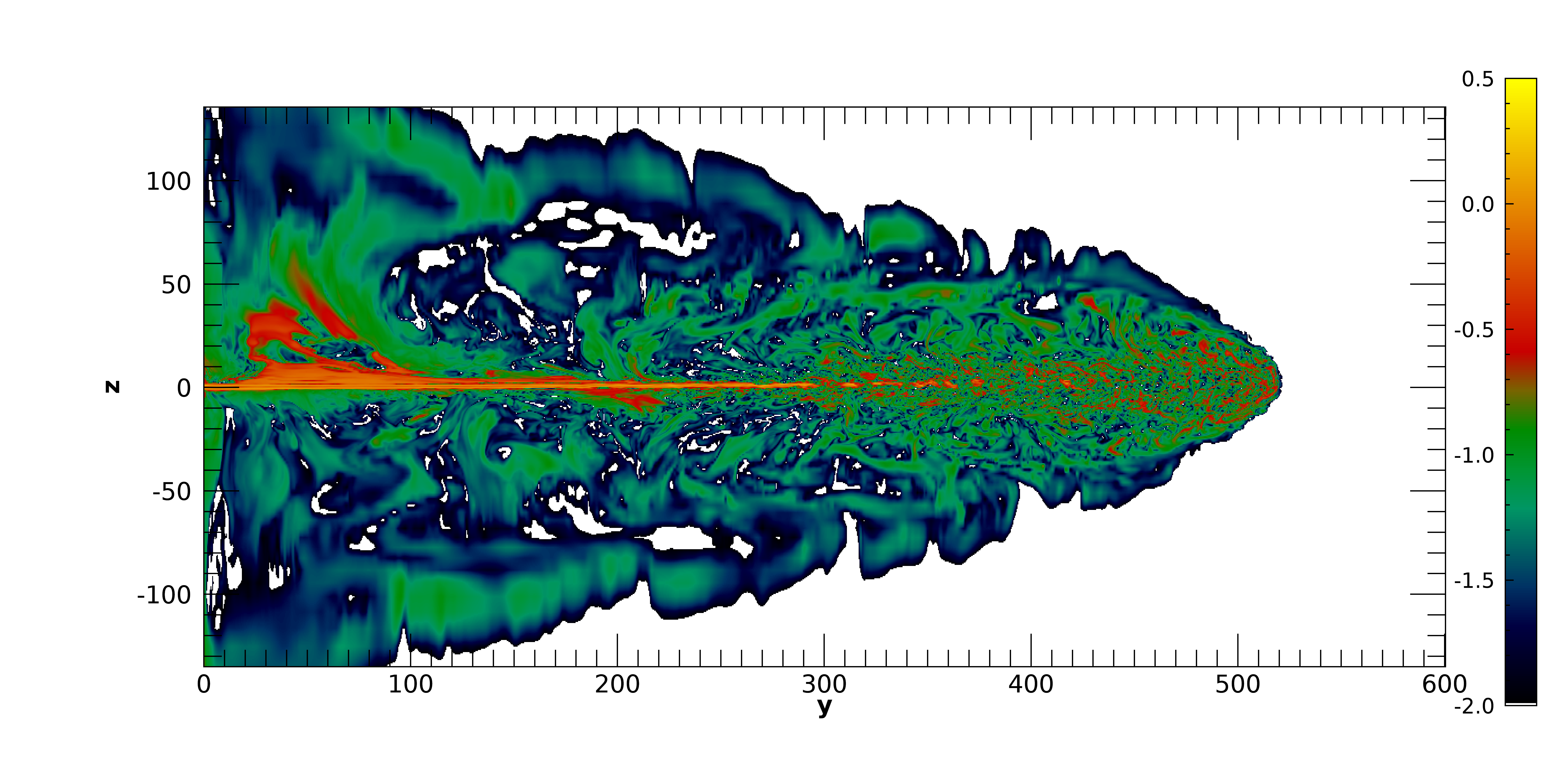}
    \includegraphics[width=\columnwidth]{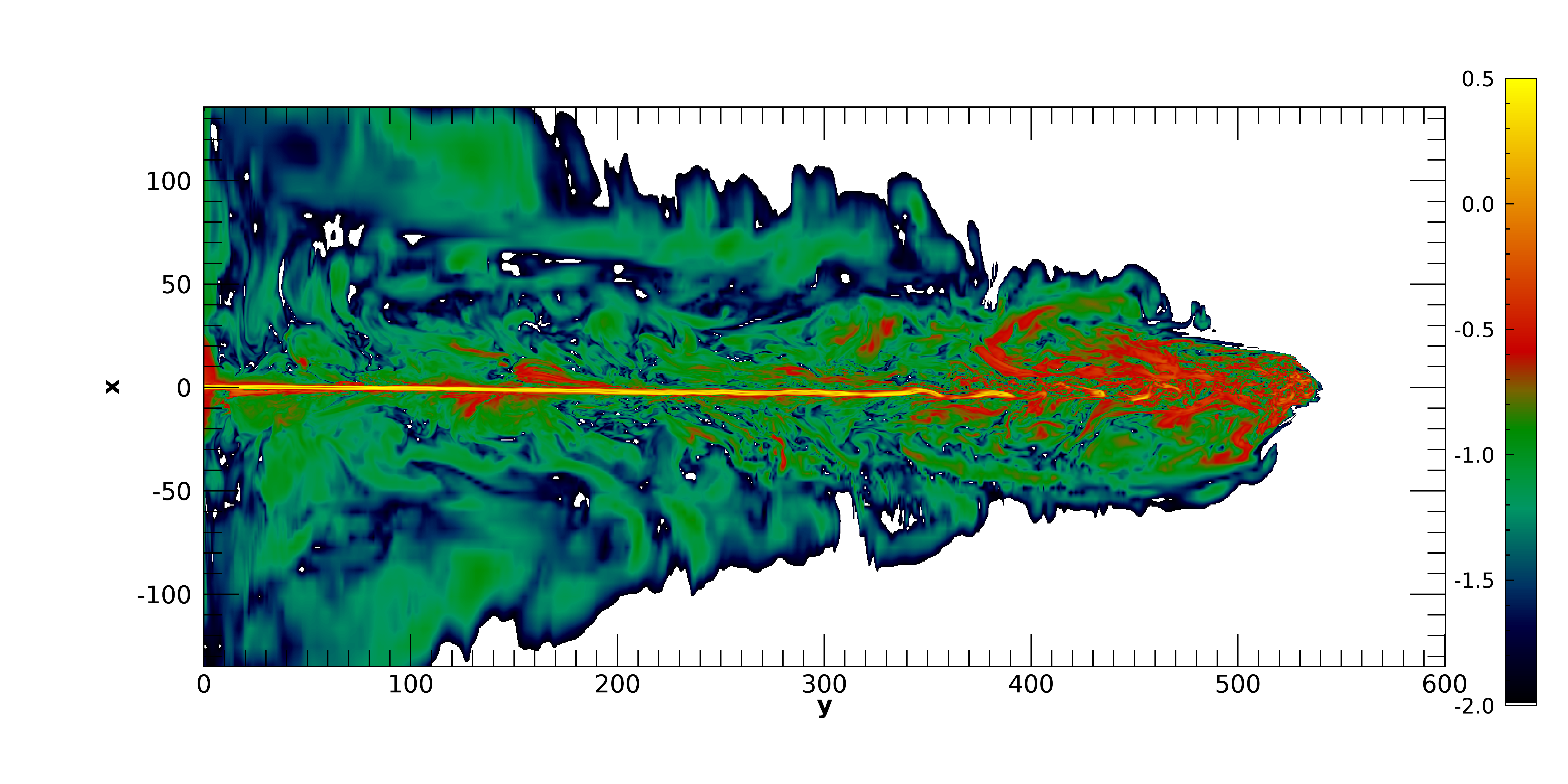}

    \caption{Longitudinal cut (in the $y-z$ plane) of the distribution of the magnetic field strength at the final times of the simulations. The four panels refer to the four different cases (Case A top-left, Case B top-right, Case C bottom-left,  Case D bottom-right).  The times are the same as in Fig. \ref{fig:gamma3D}.}
    \label{fig:bfield}
\end{figure*}

In the entrainment process, the jet transfers its momentum to the external medium. A quantitative view of this process is provided by Fig. \ref{fig:momfrac} where we plot, as a function of time, the maximum distance at which a given fraction of the momentum flux is carried by material moving at $\gamma \geq 5$. We start our discussion by looking at the top left panel, corresponding to case A. In the figure we see several curves: starting from the top, the blue curve represents the head position, the region between the top curve and the second curve refers to a jet section in which the high-velocity material carries a fraction of momentum flux $< 10\%$; the second region refers to a fraction $10\% - 20\%$ and so on; the bottom curve refers to a fraction of $70\%$. Up to $t \sim 3000$ all the curves follow the behavior of the jet head, subsequently all the curves tend to flatten and for $t \gtrsim 8000$  they become, on average, flat. As discussed in Paper II, the  flattening shows that the jet has reached an almost quasi steady state in which it progressively entrains external material and slows down. We can see that, in this quasi steady state, the jet up to $y \sim 250$ propagates almost undisturbed, with the high velocity material carrying most of its momentum, in the region $250 \lessapprox y \lessapprox 400$ the jet progressively transfer its momentum to the entrained external material, the momentum fraction carried by the high velocity material drops from $70 \%$ to $< 10 \%$, finally in the region $y > 400$ almost all the initial jet momentum is carried by slow moving material. We can notice that, as discussed above,  the region between $y \sim 250$ and $y \sim 400$ corresponds to the region where the perturbations progressively increase their amplitude until the jet becomes fragmented.   In the bottom-left panel we show the results for case C, which has a density contrast of $\eta = 10^{-5}$ and the same magnetization, $\sigma = 10^{-2}$ of case A. In this case, a quasi steady state is almost reached, all the curves tend to flatten, but, on average, they are still increasing. The process of entrainment and deceleration occurs at a larger distance with respect to case A, i.e. from $y \sim 350$ to $y \sim 450$ and can be connected, as in case A,  with the growth of perturbations and jet fragmentation that is observed in Fig. \ref{fig:gcut}.

In cases C and D, that have higher magnetization ($\sigma = 0.1$), the behavior is quite different. We already saw that the entrainment process is much less efficient with respect to the other two cases and therefore the jet does not lose its momentum almost up to its head. Without entrainment and transfer of momentum to the external material we would not observe any deceleration and, in this case, we would expect all the curves to be coincident with the jet propagation. This is confirmed by the behavior of the different curves in the right panels of Fig. \ref{fig:momfrac}  (top, Case B, $\eta = 10^{-4}$; bottom, case D, $\eta = 10^{-5}$), that follow closely the blue line representing the position of the jet head.

Combining the information presented above, we derive a different scenario for the low magnetization cases A and C with respect to the high magnetization case B and D. In cases A and C, the jet is subject to Kelvin-Helmholtz instability, that drives turbulent entrainment,  and progressively releases its momentum into the ambient medium; it widens its cross-section maintaining a high gamma value only in a central spine until it breaks and fragments into high-velocity blobs, surrounded by an envelope at $\gamma \leq 2$. In case A, however, this process is more efficient and, in the last section of the jet, the high velocity material is almost absent. In case C, the process is less efficient it occurs later  and fragments of high velocity material are often present in the last section of the jet, as  we will show later. The cases B and D, with higher magnetization,  appear to be less prone to Kelvin-Helmholtz instabilities, but the jet presents larger scale wiggles that are most likely due to current-driven instabilities. These differences between low and high magnetization cases have also been already discussed by \citet{Mig2010, massaglia19, mukherjee20} and \citet{massaglia22}. In these papers it has been shown how the larger toroidal magnetic fields stabilize the jet against short wavelength Kelvin-Helmholtz modes that are responsible of the turbulent entrainment, but lead to  current-driven instabilities that result in larger scale jet wiggling  \citep[see also][]{lopez-miralles22}, in the formation of larger scale fragments, and in a lower efficiency of entrainment. The development of current-driven instability is of course enhanced by the presence of only the azimuthal component of magnetic field. In a more general field configuration, the behavior can be somewhat different.

\subsection{Jet deceleration}
The entrainment process discussed above may lead to different connected consequences for the jet dynamics.
First, we can have a decrease in the strength of the terminal shock up to its disappearance. In \citet{massaglia16} the authors interpret this disappearance as a signature of the transition from FR~II to FR~I; in fact the presence of a strong terminal shock can be connected with the presence of
hot spots typical of FR~II radio sources. As a second point, the Lorentz factor in the terminal part of the jet, i.e. in the region close to its head, may tend to decrease, and, finally, the jet head velocity also decreases. In Figs. \ref{fig:compositeA}-\ref{fig:compositeD} we give representations of these different aspects for the four different cases. More precisely, in the top-left panel of each figure we show a scatter plot of the maximum pressure in the computational domain, as a function of time (bottom $x$-axis) and jet head position (top $x$-axis), the coloured band highlights the range of variation and the general trend. In the top-right panel we plot the $y$ position of the maximum pressure, again  as a function of time (bottom $x$-axis) and jet head position (top $x$-axis). In the bottom-left panel we plot the maximum value of the Lorentz $\gamma$ near the jet head as a function of time (bottom $x$-axis) and jet head position (top $x$-axis). More precisely, in the figure, we plot the maximum value of $\gamma$ found in the region $y_{h} -50 < y < y_{h}$, where $y_{h}$ is the $y$ coordinate of the position of the jet head. Finally, in the bottom-right panel we show the velocity of the jet head again as a function of time (lower axis) and of the head position (upper axis).

In case A (Fig. \ref{fig:compositeA}), in the top-left panel, we can observe the presence of two phases, at first the maximum pressure shows a decreasing trend followed by a flatter behavior, that starts at  $t \gtrsim 10^4$, when the jet head has reached $y \sim 500$.  In the flat part, the maximum pressure remains typically at a low level, with some sporadic spikes. Correspondingly, in the top-right panel, we observe that, up to $t \sim 10^4$ and $y_h \sim 500$, the maximum pressure is found at the jet head, while, for larger times, there are intervals in which the maximum pressure is observed in the first section of the jet. This indicates that the terminal shock has disappeared and thus the jet has acquired a morphology more similar to an FR~I. We can notice that the time at which there is the occurrence of the transition between the two phases corresponds to the time at which we observe, in the top-left panel, the transition to a quasi steady-state. The same two phases can also be observed in the bottom panels. In the bottom-left panel, we can see that, for $t < 12000$ ($y_h < 500$), on average, the maximum value of $\gamma$ is around $8$, for larger times, $t > 12000$ ($y_h > 500$), it drops abruptly to a value of about 2, with some isolated peaks, which are anyway $< 4$, except for very few instances. The jet deceleration is confirmed by the bottom-right panel, where we observe, in the first phase, strong oscillations of the jet head velocity, a smoother behavior is instead present in the second phase.  As we discussed earlier, the time of transition between the two phases corresponds to the time at which the jet reaches an almost quasi steady state. Overall,  in case A, that shows the more pronounced effects of entrainment, the jet head velocity decreases, from the beginning to the end, by a factor slightly larger than 10.

For case C (Fig. \ref{fig:compositeC}), in the top  left panel, we similarly observe a two-phase behavior: in the first phase, up to $t \sim 25000 $ and $y_h \sim 420$, we observe  a steady decrease, followed again by a flatter part. The range of variation is however larger than in case A and   high pressure peaks  occur much more frequently. Correspondingly, in Fig. \ref{fig:compositeC}, top-right panel, we observe that the intervals in which the terminal shock disappears are very sporadic, therefore the jet has almost always a FR~II-like morphology. Again the transition between the two phases occurs at the time when the jet has almost reached a steady state. Similarly, in the bottom-left panel we observe that for $t < 25000$ ($y_h < 400$)  the average  value of $\gamma$ is again around $8$, at larger times it shows strong oscillations with an average value around $4$, but dropping quite  often to values around 2. Also the behavior of the jet head velocity is similar to case A, with somewhat larger oscillations in the second phase. In fact, as discussed above, in the quasi steady state, we have more high velocity fragments that reach the jet head and this leads to the observed sporadic increases of the jet head velocity. The velocity decrease is somewhat lower than in Case A.

In the more magnetized cases B and D (Figs. \ref{fig:compositeB} and \ref{fig:compositeD}), in the top-left panels,  we do not observe the two-phase behavior, but only  a decreasing trend with a quite large interval of variation and the values of the maximum pressure remains typically higher than in cases A and C.  The position of the maximum pressure, shown in the top-right panels, is always at the jet head (with only very few exception for case D) and the maximum values of $\gamma$ stays, for all times, around $8$, with some sporadic drops at lower values, particularly for case D. The jet head velocity, in the bottom-right panels, shows a strongly oscillating behavior for all  times. The oscillations can be connected to the swirlings of the jet head induced by the current-driven kink instability. This is demonstrated by Fig. \ref{fig:kink}, where we show a sequence of three-dimensional views of a tracer iso-contour during the interval of time marked by the two black vertical lines in the bottom-right panel of Fig. \ref{fig:compositeB} (Case B). We can see that, in correspondence to the decrease of the jet head velocity, the jet head is deflected at almost $90$ degrees and then the jet returns to propagate straight and the velocity to increase again. Overall, as a result of the lower efficiency of the entrainment process, the jet head velocity for these cases has a lower decrease with respect to the previous ones. 

The structure of the magnetic field appears also quite different between low and high magnetization cases, as shown in Fig. \ref{fig:bfield}, where we display longitudinal cuts, in the $y-z$ plane, of the distribution of the magnetic field intensity. In the two left panels, corresponding to the low magnetization cases,  we can see that for $y \geq 300$ the magnetic field is characterized by small scale structures. This occurs in the region where turbulent entrainment is most effective.  At the opposite, higher magnetization cases in the right panels show a much more coherent structure of the magnetic field concentrated along the jet for almost all its length. This is consistent with the results of \citet{mukherjee20} where similar differences in the magnetic field distribution are found between the cases dominated by  the Kelvin-Helmholtz instability (as our cases A and C) and the cases dominated by the current-driven kink instability (as our cases B and D). 

\section{Summary and conclusions}
\label{sec:discussion}
We presented three-dimensional simulations of relativistic magnetized jets considering four cases with the same Lorentz factor $\gamma = 10$, but different density ratios and different magnetization, for studying the deceleration process in connection with the FR~I-FR~II dichotomy.  We follow the propagation  up to about 600 pc within the galaxy core, so our assumption of constant density is well justified. Since in FR~I radio sources the jets are relativistic at their base, but they become sub-relativistic farther out, our simulations aim, as in Papers I and II, at investigating under which conditions highly relativistic collimated flows can decelerate to sub-relativistic velocities within a distance of about one kiloparsec. Here we extend the analysis performed in Papers I  and II, by considering the effects of magnetic field.

Different kinds of instabilities may play a role in shaping the dynamics during the jet propagation: in particular Kelvin-Helmholtz instability, as shown in Papers I and II, can lead to mixing, transfer of momentum and then jet deceleration. The presence of magnetic field, on the other hand, may have a strong impact on  the instability evolution, in fact it can stabilize the jet against Kelvin-Helmholtz instabilities, but it can promote the growth of current-driven instabilities, leading to strong jet bendings and deflections \citep[see e.g.][]{Mig2010}.

 In Paper I and II we showed how lighter jets,  are more prone to deceleration. In particular the most promising case was Case C of that paper with a density ratio $\eta = 10^{-4}$. In the present paper we then consider cases with parameters similar to Case C, with $\eta = 10^{-4}$, in addition we also considered cases with a lower density ratio ($\eta = 10^{-5}$). For both values of the density ratios we considered different magnetization. Our results show striking differences between low and high magnetization cases: while the low magnetization cases (A and C) show the effects of entrainment and momentum transfer to the external medium, in high magnetization cases (B and D) entrainment and momentum transfer appear to be  much less efficient. As discussed above, this difference is due to the magnetic field that has a stabilizing effect on  Kelvin-Helmholtz instabilities and reduces the mixing. On the other hand, stronger magnetic fields can drive current-driven instabilities and, in fact, in  cases B and D we observe larger scale wigglings and jet bending. Comparing Cases A and C, we can observe that in case C the entrainment process is somewhat less efficient than in case A and also it starts at a more advanced position along the jet. The starting of entrainment in case A appears to be connected with the presence of strong recollimation shock at the jet base. The presence of recollimation shocks can favour the growth of instabilities \citep{costa23, bromberg23, gourgouliatos18a, gourgouliatos18b}, in fact, after the shock both the Lorentz factor and the Mach number of the flow  are reduced, making the conditions more favourable for Kelvin-Helmholtz instabilities. In addition, one may also observe centrifugal instabilities, recently discussed by \citet{gourgouliatos18a}.  Case C, at the opposite presents only much weaker recollimation shocks, this difference may be related to the much higher jet pressure in Case A.  In this context, with respect to the more magnetized cases B and D, we can note that \citet{mizuno15}, \citet{marti16} and \citet{marti21} have shown that stronger magnetic field can avoid the development of strong recollimation shocks, this effect may also contribute to the increased stability of these cases with respect to KH modes.  

In Paper II we showed that low density and therefore low power jets can be effectively decelerated from highly relativistic to sub-relativistic velocities within the first kpc. Then they can, on larger scales, display morphologies typical of FR~I radio sources \citep{massaglia16}.  In this paper we conclude that the low magnetization cases behave in a way much similar to the pure hydrodynamic cases discussed in Paper II. More precisely, these jets reach a quasi steady state in which we observe the progressive disappearance, along the  jet, of material moving at high Lorentz factor. In the last section of the jet we have a broad flow moving at $\gamma < 2$ interspersed with blobs at higher Lorentz factor. In case A, we observe also the disappearance of the terminal shock, and therefore, from the observational point of view, of the hot spot. The jet in this case loses the typical characteristic of a FR~II radio source and acquires,  already at these early stages of the jet evolution,  an appearance that can be connected to FR~I radio-sources. As discussed by \citet{capetti20}, the presence of small scale and/or low brightness jets in FR~0s indicates that FR~0s and FR~Is can be interpreted as two extremes of a continuous population of radio sources, characterized by a broad distribution of sizes and luminosities of their extended radio emission. Depending on the physical scale at which the jet disruption occurs, Case A can 
be associated with both classes (FR~0 and FR~I) of radio sources with an edge-darkened morphology. 

At the opposite, the higher magnetization cases B and D, maintain the jet flow well collimated up to the end of our simulations. The jets show some fragmentation, but contrary to the previous cases the fragments are quite large. In addition, high Lorentz factors are maintained up to the jet head and we always observe the presence of the terminal shock.  Increasing the magnetic field strength seems to make the transition to an FR~I morphology more difficult even for lower power jets. We have however to consider that we injected the jets with a pure toroidal field, the presence of a longitudinal component may somewhat change the behavior. The propagation of jets with a more complex structure of the magnetic field will be examined in a future work.  Other aspects may as well play a role in the jet entrainment and deceleration process, one of them is, for example the presence of an inhomogeneous external medium discussed by \citet{wagner11, wagner12}, and \citet{mukherjee16}, that may lead to a change in the interaction between the jet and the ambient medium. This will be also examined in future investigations.

\begin{acknowledgements}
 This work was supported by PRIN MUR 2022 (grant n. 2022C9TNNX)  and by the INAF Theory Grant {\it Multi scale simulations of relativistic jets}.  We acknowledge  the INAF-CINECA {\it Accordo Quadro MoU per lo svolgimento di attività congiunta di ricerca Nuove frontiere in Astrofisica: HPC e Data Exploration di nuova generazione} for the availability of computing resources. We thank M. Perucho for helpful comments.
 \end{acknowledgements}

\bibliographystyle{aa}
\bibliography{main.bib}

\label{lastpage}
\end{document}